\documentclass[sigconf]{acmart}

\usepackage{algorithm, algorithmic}
\usepackage{graphicx}
\usepackage{arydshln}
\usepackage{xurl}
\usepackage{xcolor}
\usepackage{amsmath}

\usepackage{CJKutf8}
\usepackage{multirow}
\usepackage{url}
\usepackage{amsmath}
\usepackage{subcaption} 

\AtBeginDocument{%
 \providecommand\BibTeX{{%
 \normalfont B\kern-0.5em{\scshape i\kern-0.25em b}\kern-0.8em\TeX}}}

\setcopyright{acmcopyright}

\copyrightyear{2023} 
\acmYear{2023} 
\setcopyright{acmlicensed}\acmConference[WebSci '23]{15th ACM Web Science Conference 2023}{April 30-May 1, 2023}{Evanston, TX, USA}
\acmBooktitle{15th ACM Web Science Conference 2023 (WebSci '23), April 30-May 1, 2023, Evanston, TX, USA}
\acmPrice{15.00}
\acmDOI{10.1145/3578503.3583630}
\acmISBN{979-8-4007-0089-7/23/04}

\begin{document}

\title{The Impact of Data Persistence Bias on Social Media Studies}

\author{Tuğrulcan Elmas}
\affiliation{%
 \institution{EPFL}
 \city{Lausanne}
 \country{Switzerland}}
\email{tugrulcan.elmas@epfl.ch}

\renewcommand{\shortauthors}{Tuğrulcan Elmas}


\begin{abstract}

Social media studies often collect data retrospectively to analyze public opinion. Social media data may decay over time and such decay may prevent the collection of the complete dataset. As a result, the collected dataset may differ from the complete dataset and the study may suffer from data persistence bias. Past research suggests that the datasets collected retrospectively are largely representative of the original dataset in terms of textual content. However, no study analyzed the impact of data persistence bias on social media studies such as those focusing on controversial topics. In this study, we analyze the data persistence and the bias it introduces on the datasets of three types: controversial topics, trending topics, and framing of issues. We report which topics are more likely to suffer from data persistence among these datasets. We quantify the data persistence bias using the change in political orientation, the presence of potentially harmful content and topics as measures. We found that controversial datasets are more likely to suffer from data persistence and they lean towards the political left upon recollection. The turnout of the data that contain potentially harmful content is significantly lower on non-controversial datasets. Overall, we found that the topics promoted by right-aligned users are more likely to suffer from data persistence. Account suspensions are the primary factor contributing to data removals, if not the only one. Our results emphasize the importance of accounting for the data persistence bias by collecting the data in real time when the dataset employed is vulnerable to data persistence bias. 

\end{abstract}




\maketitle
\newcommand{\Secref}[1]{Section~\ref{#1}}
\newcommand{\Figref}[1]{Fig.~\ref{#1}}
\section{Introduction}

Many data science studies employ social media data to analyze human behavior and social phenomena. They usually collect the data retrospectively, i.e. after some time since the data appears on the platform. In this case, the solicited posts that are removed \emph{before} the time of collection will not be available to the study. For instance, Twitter provides an API to academics so that they can retrieve tweets using keywords from Twitter without any time limitation. However, academics still cannot retrieve the tweets that are removed from the platform. Furthermore, some studies provide their data for reproduction. As social media platforms often disallow sharing data directly, these studies only share the unique identifiers of the data instead of the whole content, e.g., they only share tweet and user ids of Twitter data. The successor studies who reproduce them will not be able to retrieve the data that do not exist on the platform if they opt to collect the data retrospectively using the ids. In both cases, the data that is collected retrospectively may suffer from \emph{data persistence}, i.e. the data that is supposed to exist on the system may be no longer available. More importantly, the collected data may differ significantly from the original dataset that is the intended focus of the analysis. Studies that analyze controversial debates that are vulnerable to manipulation may miss out on the posts by users who were the source of the debate, who manipulated it (i.e., injected potentially harmful content), and who authored the key narratives (i.e., the topics the researchers intend to analyze). Therefore, it is crucial to investigate the data persistence for such controversial topics to see if reliable analyses are possible when the data is collected retrospectively. 

Past research suggests that the Twitter datasets suffer from data persistence, but the remaining data is still representative of the original dataset in terms of textual content~\cite{zubiaga2018longitudinal}. However, in this paper, we adopt a new approach and study \emph{data persistence bias}. Data persistence bias refers to the phenomenon where certain types of data are overrepresented or underrepresented in a dataset due to some factors causing their removal or persistence. This can lead to biased conclusions or incorrect inferences if the dataset is used for research or analysis purposes. We analyze data persistence bias on Twitter with a special focus on controversial topics that are previously analyzed in social media studies. Our contribution is to show which topics data persistence bias affects the most and quantify its impact for the first time to the best of our knowledge. To do that, we tackle the following research questions:

\begin{enumerate}
    \item \textbf{RQ1:} What is the data turnout for the datasets used by social media studies? Is it sufficient for a reliable analysis?
    \item \textbf{RQ2:} How does the data persistence bias impact social media studies, in terms of political orientation, the presence of potentially harmful content, and the source of the topic?
    \item \textbf{RQ3:} Which topics and types of content are more likely to suffer from data persistence?
    \item \textbf{RQ4:} What are the factors in data removals that may introduce data persistence bias?
    
\end{enumerate}

To answer these questions, we conducted three case studies that employ different types of datasets: 1) controversial topics, 2) trending topics, and 3) frames in the migration debate. In each case, we recollect the datasets and compute the \emph{data turnout} which is the percentage of data that was still collectible at the time of collection, after the data was created. We then analyze the bias using multiple measures which are political orientation, presence of potentially harmful content, and topics. We chose these measures as they are commonly used in social media studies. We found that the data turnout is significantly low and the datasets are biased towards the political left in the context of U.S. politics for controversial topics. Topics related to political manipulation and that are negative towards immigrants are more likely to suffer from data persistence. We also observe that the data turnout of potentially harmful content is in line with the overall data turnout for controversial topics, but is significantly low for non-controversial topics. Finally, our analysis reveals that account suspensions are the primary contributor to data removals leading to data persistence bias. However, the deletion of data by their authors is also a significant factor. Our results emphasize the importance of collecting the data in real-time over retrospectively and encourage data sharing. All code, results, and the tweet IDs are made available in \url{https://github.com/tugrulz/DataPersistenceBias}.

\section{Background and Related Work}
\label{sec:related}

\subsection{Data Persistence}
Data persistence is the availability of the data on the system. On social media, the data on the platform may be removed or made uncollectable as the platforms suspend users or users themselves remove their data. As a result, social media datasets may decay over time. Liu et al.~\cite{liu2014tweets} estimated that 20\% of tweets became unavailable in five years while Pfeffer et al. reported that the tweet decay can be as low as 45\% after 5 years~\cite{pfeffer2022sample}. Such decay may jeopardize the integrity of the social media studies that rely on them for analysis. This is because the missing data may be significantly different from the data that still exists~\cite{almuhimedi2013tweets}. Crucially, the missing data may be the focus of the study. For instance, King et al.~\cite{king2013censorship} showed that Chinese censors remove social media posts that promote collective action within 24 hours, making it impossible to study civil unrest through social media. In some cases, the data removal is the means to manipulate the social media ~\cite{torres2022manipulating} and its removal is an attempt to avoid investigation~\cite{elmas2021ephemeral}. For instance, political trolls often delete all their tweets and reset their profiles to hide their malicious activity and use their accounts in another context~\cite{zannettou2019disinformation,elmas2020misleading}. Such accounts remove the data of the old persona, making the data impossible to analyze retrospectively. Changes in platform policies also jeopardize the data persistence of potentially harmful content that researchers may want to study. For instance, Elmas et al.~\cite{elmas2021dataset} found that Twitter removed 44\% of the content that was censored but was not removed from the platform initially, due to policy changes related to hate speech. Although the removal is beneficial to the public, it denies researchers from studying such content. These findings emphasize that the retrospective collection of social media data may be problematic as it fall short of acquiring complete datasets. Yet, many studies still collect the data retrospectively through official APIs and suffer from data persistence. Therefore, it's crucial to study and understand the biases occurring due to this practice in a systematic way, which motivates our study.


To the best of our knowledge, the only similar work to ours is by Zubiaga~\cite{zubiaga2018longitudinal} who evaluated the completeness and representativeness of Twitter datasets collected in the past. They found that the number of remaining tweets can drop as low as 65\% for datasets that are 4 years old but the data turnout is 80\% on average, which is high. However, they report only on 30 manually selected event datasets. Some of these events such as the Superbowl are not controversial and thus, violations of Twitter policies are less likely, resulting in a higher data persistence. Others such as the Hong Kong Protests in 2012 might be controversial but it is from the time when the information operations might be less prevalent, or the level of content moderation might be lower. Our work reports lower turnouts on controversial topics such as QAnon that are more recent. Zubiaga also argues that the textual content of the recollected dataset is still largely representative of the original dataset using textual features. However, in our work, we focus on bias, using metrics such as the political orientation. We find that for controversial datasets, the remaining data leans towards a certain political orientation, which may prevent analyzing what the other side may argue. Lastly, our work is not limited to a few datasets, and attempts to root out the datasets that are likely vulnerable to data persistence from a pool of datasets.

\subsection{Biases on Social Media Studies} Past studies pointed out possible biases when working with social media data, including data persistence~\cite{boyd2012critical, olteanu2019social}. For instance, Gaffney et al.~\cite{gaffney2018caveat} identified critical gaps in a widely used open-source Reddit dataset and pointed out potential risks using such missing data in terms of user and network analyses. However, only a few studies attempt to quantify such bias and report what they can imply for social media studies to the best of our knowledge. Yang et al.~\cite{yang2020twitter} compared tweet-level sampling with user-level sampling and argued that the former introduces a bias towards hyperactive users involved in inauthentic behavior, as they are overrepresented in the data. Wu et al~\cite{wu2020variation} focused on Twitter API sampling across different timescales and reported a significant impact on network structures and retweet cascades. Gonzalez et al~\cite{gonzalez2014assessing} also studied API biases on Twitter and found that search API overrepresents central users on the social network. All these studies focus on sampling biases. Our work focuses on data persistence bias instead, which is the first to quantify it, and analyze its impact. To do that, it employs multiple measures such as political orientation, presence of potentially harmful content and topics, which are common to social media studies such as analyzing stance~\cite{grvcar2017stance}, polarization~\cite{ozer2019measuring}, filter bubbles~\cite{elmas2020can}, elections~\cite{bovet2018validation}.

\section{Methodology}
\label{sec:methodology}

We now present the methods we used for one or multiple case studies. We leave the case-specific details to their respective sections.

\smallskip
\noindent\textbf{Dataset Creation:} As our study focuses on data persistence, it requires retrospective datasets of tweets that include deleted tweets. To collect such dataset(s), we employ the Internet Archive's Twitter Stream Grab, which has been used extensively by past research~\cite{tekumalla2020mining,elmas2022characterizing}. It is collected through Twitter's official Spritzer API, which provides the 1\% random sample of all tweets. The dataset contains the tweets between 2011-2021 and is publicly available~\cite{team2020archive}. Crucially, in this dataset, the tweets are collected \emph{as} they are posted (i.e. real time) and the tweets that are later deleted are not removed. We use this dataset to create new datasets in Case 1 and Case 2. We also create a \textit{control dataset} out of this dataset, by randomly sampling 10,000 tweets per day in 2020 (except for the 17 days for which the Internet Archive did not archive the data). It consists of 3,480,000 tweets. We recollect the datasets using Twitter API in October 2021. We then compute the \emph{tweet turnout} for each dataset by dividing the number of tweets we could collect by the number of tweets in the original dataset. We also compute the \emph{author turnout}, which is the number of authors in the recollected dataset divided by the number of authors in the original dataset.

\begin{figure}[!htb]
    \centering
    \includegraphics[width= .92\columnwidth]{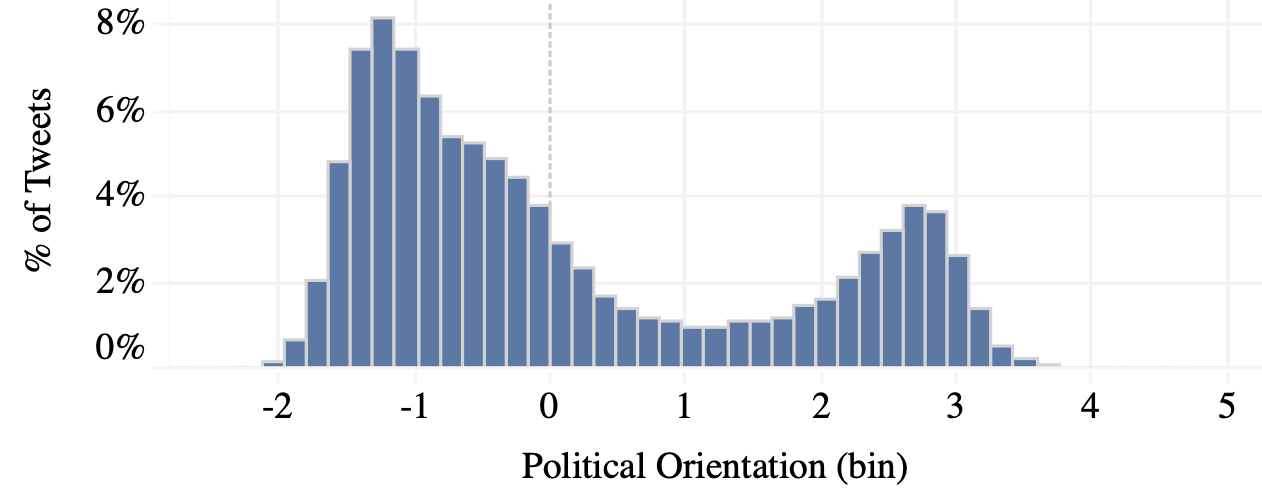}
    \caption{The distribution of political orientation scores for the control dataset.}
    \label{fig:reference_scores}
\end{figure}

\smallskip
\noindent\textbf{Political Orientation:} We quantify bias using the change in political orientation within the datasets as one of the measures. We do this in the context of U.S. politics, so we classify each user by their leaning towards Democrats or Republicans. We reproduce the study by Barbera~\cite{barbera2015birds}. Their methodology employs Bayesian inference and uses follower data to assign a political orientation score to each user. The scores are between -5 to +5. Positive values signify leaning toward Republicans and negatives signify leaning toward Democrats. The scores do not follow a normal distribution. As ~\Figref{fig:reference_scores} shows, the majority (61\%) of the authors of the tweets in the control dataset have political orientation score less than 0, i.e. classified as leaning toward the left. Additionally, 95\% of users have score between -1.6 and 3.2. These findings have two implications. First, the scores depicting leaning toward the left are denser: up to 8\% of left leaning users have a political orientation score difference by 0.1. Second, although the distribution of scores depicting the right are wider, the right-leaning users are in minority. This means that even if the political orientation score of a dataset is shifted towards the left upon recollection by a small margin, it prevents analysis of many users that are aligned with right which were the minority in the first place.

Our caveat with this political orientation classification method is that not all users are assigned a score. We discard such users when we report biases measured by political orientation. We assume that the methodology has sufficient coverage for the analysis, i.e. we have scores for a sufficient number of users to reliably assess the bias. This is because the methodology relies on following users with known political orientations. This signal is less sparse than others such as tweeting or retweeting that are employed by other methods. Thus, the methodology may be more likely to reach the theoretical maximum of users for which we can infer the political orientation of, or the users with any political orientation. For reference, we have political orientation scores for the 15\% of tweets in the control dataset. 

\smallskip
\noindent\textbf{Potentially Harmful Content:} We also study bias in terms of the presence of potentially harmful content. We detect such content by using tweets with negative sentiments and hate speech as proxies. For the former, we use VADER, a widely used rule-based model for general sentiment analysis~\cite{hutto2014vader}. It returns sentiment scores between -1 and 1. Negative values indicate tweets with negative sentiments. We used a threshold of -0.5 to classify tweets as tweets with negative sentiments. To detect hate speech within the text, we used Jigsaw and Google's Perspective API~\cite{Perspective}. It is a public API that assigns hate scores based on a transformers model~\cite{lees2022new}. The API returns scores between 0 and 1 that indicate the presence of different types of hate speech, which are "Toxicity", "Severe Toxicity," "Insult", "Threat", "Profanity", and "Identity". For simplicity, we assign the maximum score to each tweet and used a threshold to indicate that the tweet contains any kind of hate speech. We chose 0.61 as the threshold which was suggested in~\cite{kumar2021designing}.

\smallskip
\noindent\textbf{Topic Analysis:} To analyze the topics in Case 1, we employed word shift graphs. Wordshift graphs highlight the differences between two sets of text by showing the most distinctive n-grams in each set using Shannon entropy~\cite{gallagher2021generalized}. We only used entities (hashtags and mentions) and bigrams as the input, as we observe that they are more descriptive than unigrams.

\section{Case 1: Controversial Topics} 
\begin{figure}[!htb]
    \centering
    \includegraphics[width = \columnwidth]{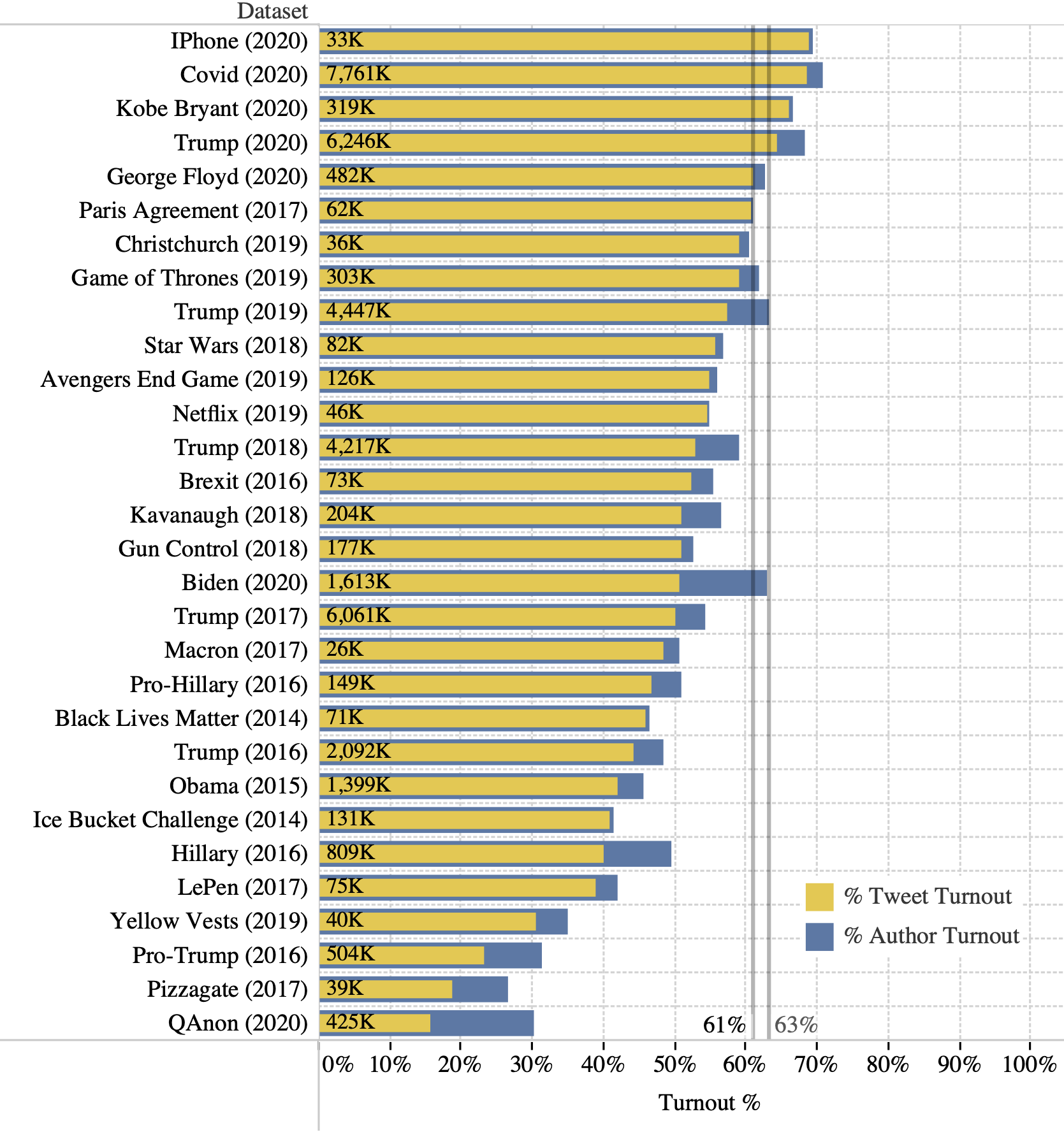}
    \caption{The tweet and author turnouts of the datasets. The labels over the bars indicate the number of total tweets. The plot is sorted by the tweet turnout. The vertical lines show the tweet and the author turnout of the control dataset.}
    \label{fig:turnout_data}
\end{figure}


\noindent\textbf{Datasets} Many social media studies focus on controversial topics related to important political events. They analyze public opinion to understand the arguments and counter-arguments, and study group formation within the datasets. The data persistence bias can alter the results of such studies as they would alter the groups, and under-represent some of the opinions and arguments. To test this, we replicate those studies. We choose several social media studies analyzing controversial topics and simulate their data collection strategy, which is collecting data through manually selected keywords that are relevant to the topic of interest over multiple days. We collect the datasets from the 1\% sample provided using such keywords. The topics and the studies we choose are the following: QAnon, a viral conspiracy theory that led to Capitol Riot in 2020,~\cite{sipka2021comparing}, Pizzagate, another viral conspiracy theory~\cite{metaxas2019investigating}, candidates during the 2016 U.S. Presidential Election Trump, Hillary (2016)~\cite{bovet2018validation}, American presidents, Obama, Biden, and Donald Trump, George Floyd, whose murder instigated nationwide protests in U.S. in 2020, ~\cite{toraman2022blacklivesmatter}, Paris Agreement (using "paris agreement" or "climate change") which sparked controversy when Donald Trump ordered withdrawal from it~\cite{marlow2021bots}, Gun Control (using "gun control" and "gun violence") during 2018 protests after a school shooting~\cite{ozer2019measuring}, Kavanaugh, whose nomination to supreme court instigated controversy~\cite{darwish2019quantifying}, Black Lives Matter, an anti-racist movement in U.S.~\cite{giorgi2022twitter}, and Brexit~\cite{grvcar2017stance}. We also collect datasets related to French politics, the presidential candidates in 2017 elections, Macron, and Le Pen, and the Yellow Vest movement in 2019, as those topics were targeted by the American far-right community~\cite{ferrara2017disinformation}. Additionally, we collect datasets on topics that are not controversial such as Covid, Kobe Bryant (after his death), Netflix (in the first three months of 2019), iPhone (during the release of iPhone 12), Star Wars (during the release of Episode VIII), Ice Bucket Challenge (when it was viral in 2014), Game of Thrones (during the last season in 2019), Avengers End Game (during its debut in 2019. We create each dataset by only using its name (unless we explicitly state what we also use in the parenthesis) as the query in order not to create a keyword-based bias e.g., for the Trump dataset we use only "trump" as the keyword to collect data. However, we create two additional datasets, pro-Trump and pro-Hillary, which mention the keywords and hashtags in support of one of the candidates, listed in ~\cite{bovet2018validation}. Additionally, for all keywords, we also use their hashtag equivalents, e.g. for "gun control" we also use \#guncontrol. For covid, we use the other names of the disease (e.g., sars-cov). The simulated collection period is the time period when the topic was popular. As "Trump" was popular for multiple years, we created a separate dataset for each year. 

\smallskip
\noindent\textbf{Turnout} To answer \textbf{RQ1}, we recollect the datasets and evaluate tweet and author turnout. As ~\Figref{fig:turnout_data} shows, the maximum data turnout is 70\% (iPhone) and can go as low as 15\% (QAnon). The data turnout is generally lower on controversial datasets and older datasets. We found that the datasets which represent topics that were manipulated by coordinated groups such as Pizzagate, Qanon, Yellowvests, Le Pen have very low turnout. This is likely due to Twitter's intervention. For instance, in January 2021, Twitter announced that they suspended 70,000 accounts promoting QAnon-related content~\cite{twittersafety}, which made 85\% of the relevant data inaccessible as our study shows. Additionally, the older datasets that date back to 2016 all have data turnout below 50\%, signifying that the majority of the tweets in them are uncollectable.

The turnout for the control dataset is 61\%. Only iPhone (2020), Covid (2020), Kobe Bryant (2020), Trump (2020) and George Floyd (2020) has higher turnout than the control dataset, which are all recent datasets. The difference in turnout between the control dataset and the other datasets is statistically significant according to the test for proportions based on the chi-squared test except for the Paris Agreement dataset, which has a turnout of 60.8\%. Overall, the turnout rates are generally very low even after a year which posits a problem for social media studies using retrospective data, especially when the focis is on controversial topics.

We also evaluate author turnout to see if the bias is sourced from a few hyperactive users as in the case in~\cite{yang2020twitter}. For most of the datasets, the author turnout is roughly the same as the tweet turnout, e.g., it is 63\% for the control dataset, which is only greater than the tweet turnout by 2\%. The difference is higher for the datasets which may be the target for coordinated groups that tweet aggressively (i.e. in higher frequencies) and be overrepresented within the dataset. The following datasets have more than a 3\% difference between two types of turnout and the difference is statistically significant (p < 0.0001): QAnon, Pizzagate, Yellow Vests, Hillary, Biden, Obama, Brexit, and all Trump-related datasets. Our caveat is that our dataset is based on 1\% in which users central to the topic are reported to be over-represented~\cite{gonzalez2014assessing}. Thus, we may have fewer over-active users in the complete datasets and those users may increase author turnout, which may close the gap between the tweet and author turnout. 

\begin{figure*}[!htb]
    \centering
    \includegraphics[width = 
    0.7\textwidth]{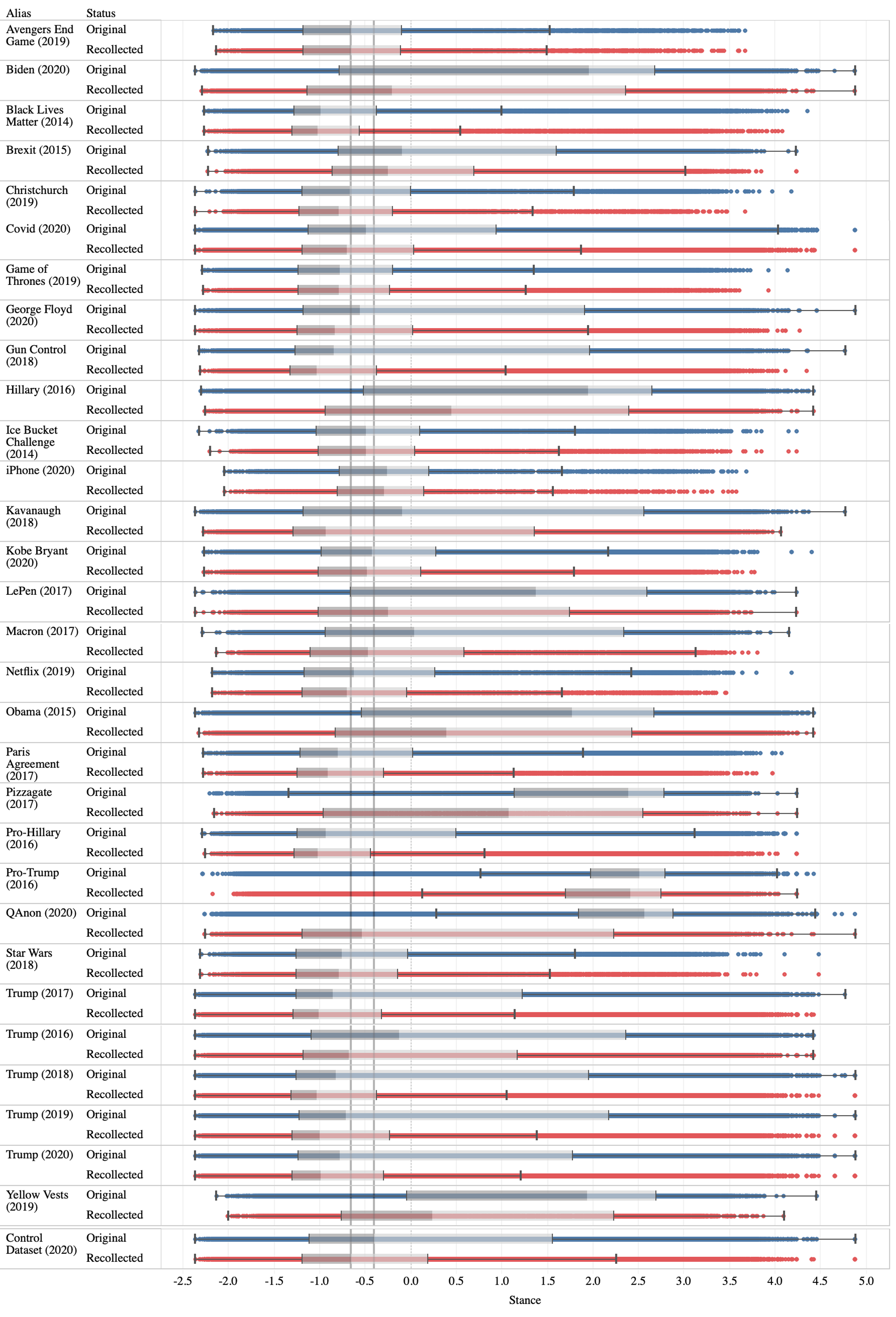}
    \caption{The political orientation distribution for the original and the recollected datasets. The vertical lines refer to the median scores for the original and the recollected control dataset.}
    \label{fig:political_orientation}
\end{figure*}

\smallskip
\noindent\textbf{Political Orientation} To answer \textbf{RQ2}, we first measure the bias by political orientation. Upon recollection, we found that almost all datasets are biased towards Democrats. The impact is higher on datasets with controversial topics but negligible for non-controversial datasets. \Figref{fig:political_orientation} shows the results. The mean political orientation score of the control dataset decreased from 0.15 to -0.23. Interestingly, the difference (0.39) is higher than the majority of the datasets. The following datasets passed this threshold and observed a higher difference: Biden (2020), George Floyd (2020), Gun Control (2018), Hillary (2016), Kavanaugh (2018), LePen (2017), Macron (2017), Obama (2015), Pizzagate (2017), QAnon (2020), Trump (all) and Yellow Vests (2019). The differences are statistically significant according to Welch's t-test (p < 0.0001).

We also observe that even though the magnitude of the change in mean scores is small, the quartiles in the political orientation distribution lean towards the left dramatically. ~\Figref{fig:political_orientation} shows the box plot of the stance distributions for each dataset. We observe that the upper quartile of the control dataset decreased from 1.5 to 0.2. This suggests that Twitter purged many polarized Republican-oriented users in 2020 which may not be studied regardless of the topic of interest. The results are even more dramatic for the conspiracy theories in our study. The lower quartile of the original Pizzagate dataset is 1.14. It is roughly equal to the median of the recollected Pizzagate dataset, 1.07. The upper quartile of the recollected QAnon dataset (2.23) is also close to the lower quartile of the original (1.84). 

\smallskip
\begin{figure*}[!htb]

    \begin{subfigure}{.45\textwidth}
        \centering
        \includegraphics[width=\linewidth]{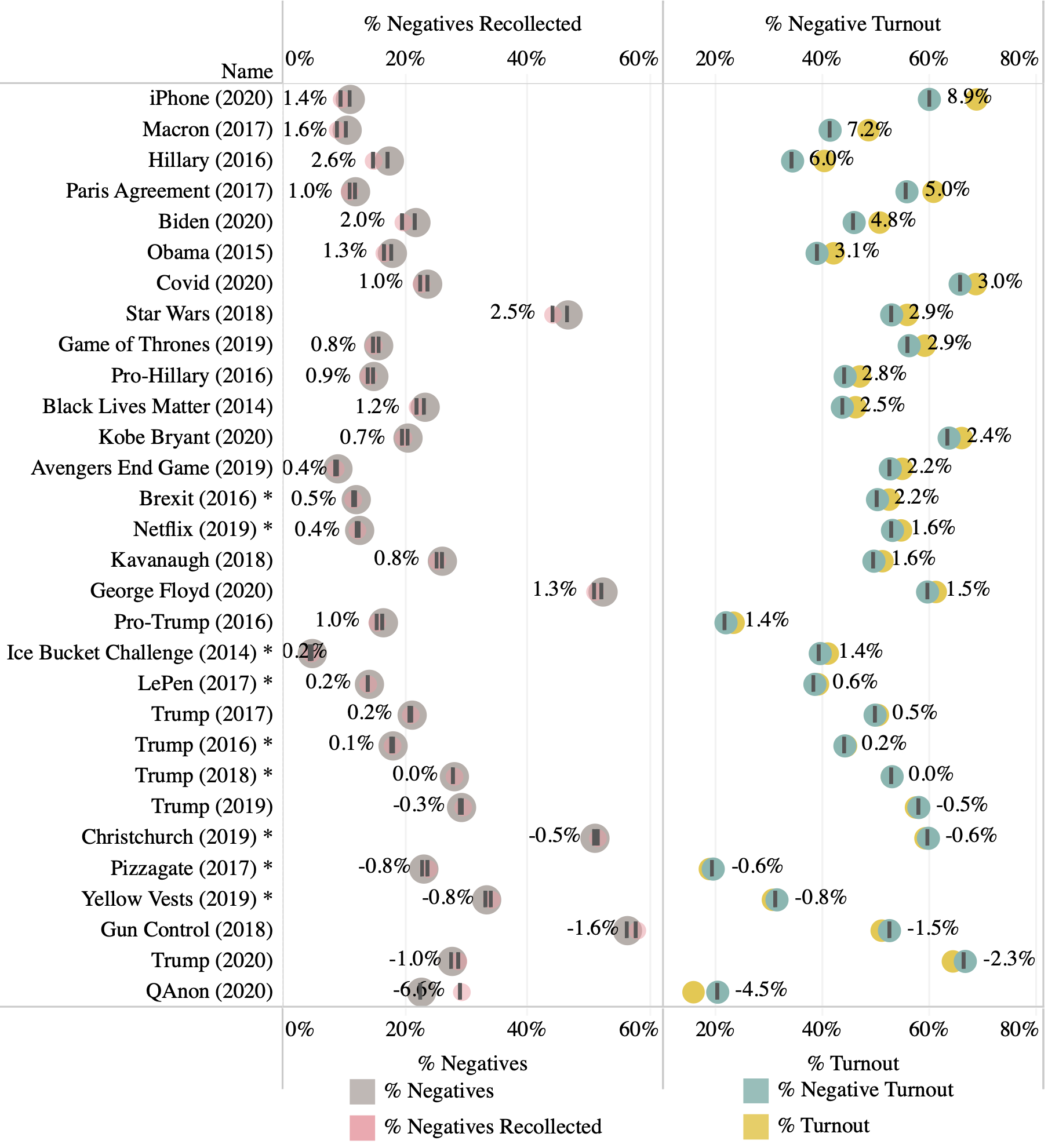}
    \end{subfigure}
    \begin{subfigure}{.45\textwidth}
        \centering
        \includegraphics[width=0.96\linewidth]{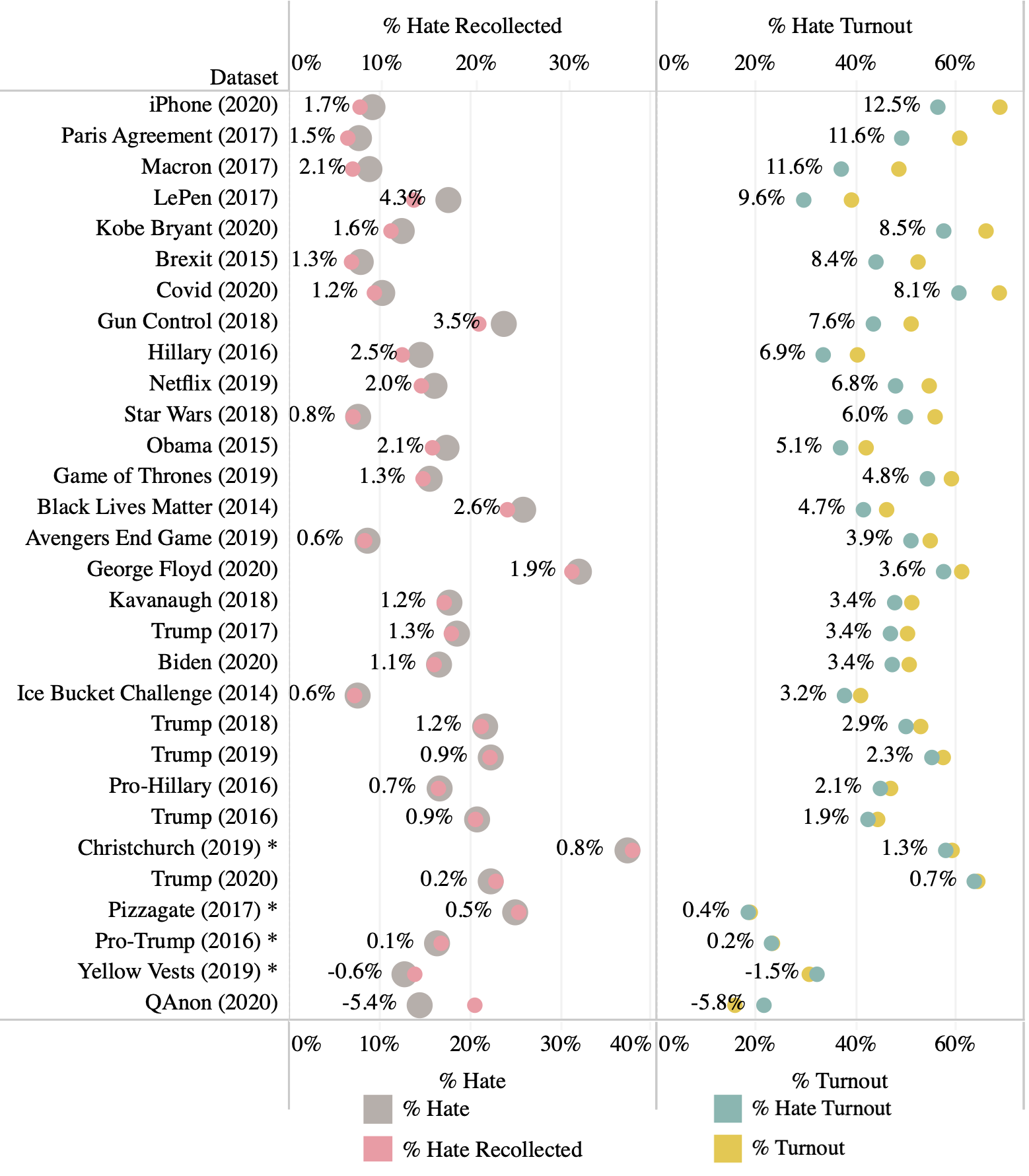}
    \end{subfigure}
    \caption{The decrease of the share of the potentially harmful content (left) and its turnout's comparison with the overall turnout (right). (*) The difference between the turnout of potentially harmful content and the overall turnout is NOT statistically significant according to the Chi-squared test (p < 0.0001).\label{fig:harmful_content}}

\end{figure*}
\noindent\textbf{Potentially Harmful Content} (\textbf{RQ2)} We compute the percentage of tweets with potentially harmful content in each dataset and their recollections, then calculate the difference. We also compute the turnout of tweets with such content and compare it with the overall turnout. \Figref{fig:harmful_content} shows the result. We observe that the percentage of tweets with harmful content decreases in almost all of the datasets, although the magnitude is rather small. The maximum decrease in the share of tweets with negative sentiment is observed in Hillary's (2016) dataset, where the percentage dropped from 17.3\% to 14.7\%, by 2.6\%. For the tweets with hate speech, the maximum decrease is by 4.3\% from 17.5\% to 13.2\%. Interestingly, in the QAnon dataset, the percentage of tweets with negative sentiment and with hate speech increased after recollection, by 6.6\% and 5.4\% respectively. This may be because Twitter purged the users promoting the QAnon (which may be posting tweets with positive sentiment) and left the users who disregard the conspiracy theory with negative sentiment. We also observe that the turnouts of the tweets with negative sentiment or hate speech are mostly in-line with the overall turnouts, i.e. the potentially harmful content may have the same probability of removal as other types of content. Nonetheless, we observe that the difference is higher on some non-controversial datasets such as iPhone (2020), Star Wars (2018), and Kobe Bryant (2020) and the results are statistically significant. For instance, the turnout of tweets in the iPhone dataset is 69\%, which is the highest among all datasets. However, the turnout of the tweets with hate speech is 56\%, less than the overall turnout by 12.5\%. In conclusion, data persistence may introduce a bias in terms of the presence of harmful content but the impact may be small for controversial datasets. 

\smallskip
\noindent\textbf{Topics} To answer \textbf{RQ3}, we performed a qualitative topic analysis on the missing tweets to better understand the content the studies may not be able to capture. To do that we, compared the content in the missing tweets with the recollected datasets using word shift graphs (See \Secref{sec:methodology}). For brevity, we only present the results with 15 datasets where we observe clear differences and only the top 20 n-grams. We observe that most of the missing tweets contain slogans against the topic. For instance, missing tweets in Biden (2020), Hillary (2016), and in Pro-Hillary (2016) appear to be sourced from users in favor of Donald Trump as the common hashtags include \#MAGA, \#Trump2020, \#NeverHillary, and \#CrookedHillary. We have a similar finding in Macron (2017) and Yellow Vests (2019) where the missing tweets are against Macron such as \#MacronDemission which means "Macron Resign", and \#MacronLeaks, a coordinated disinformation campaign~\cite{ferrara2017disinformation} or supportive of his opponent Marine Le Pen. This suggests that data persistence may affect studies analyzing such counter-groups. We observe that the missing tweets in the datasets related to Trump politics such as Kavanaugh (2018) and Paris Agreement (2018) also contain tweets that are supportive of the decisions, such as \#ConfirmKavanaughNow and \#AmericaFirst. The missing tweets in the conspiracy theory-related datasets are supportive of the theories (e.g., \#GreatAwakening, \#Pedogate) while the tweets in the recollected datasets contain the 2-gram "conspiracy-theory", suggesting that the remaining tweets are mostly critical of the theories. Interestingly, the recollected Gun Control dataset is more likely to contain the hashtag \#GunViolence, which is a more extreme version of the hashtag \#GunControl. We also find that the missing tweets in the Netflix dataset contain the 2-grams "Full-access", "Netflix-premium", and "watch-netflix." This may suggest that those tweets promoted illicit access to Netflix, or scam, and got removed. We also observed that retweets and mentions to popular accounts that are suspended are missing. For instance, @gatewaypundit was suspended due to spreading fake news, and all its tweets and retweets to its tweets are made inaccessible. Such removals may prevent researchers to analyze the influence of such accounts on the users.

\begin{figure*}[!htb]

    \begin{subfigure}{.20\textwidth}
        \centering
        \small Biden (2020)
        \includegraphics[width=\linewidth]{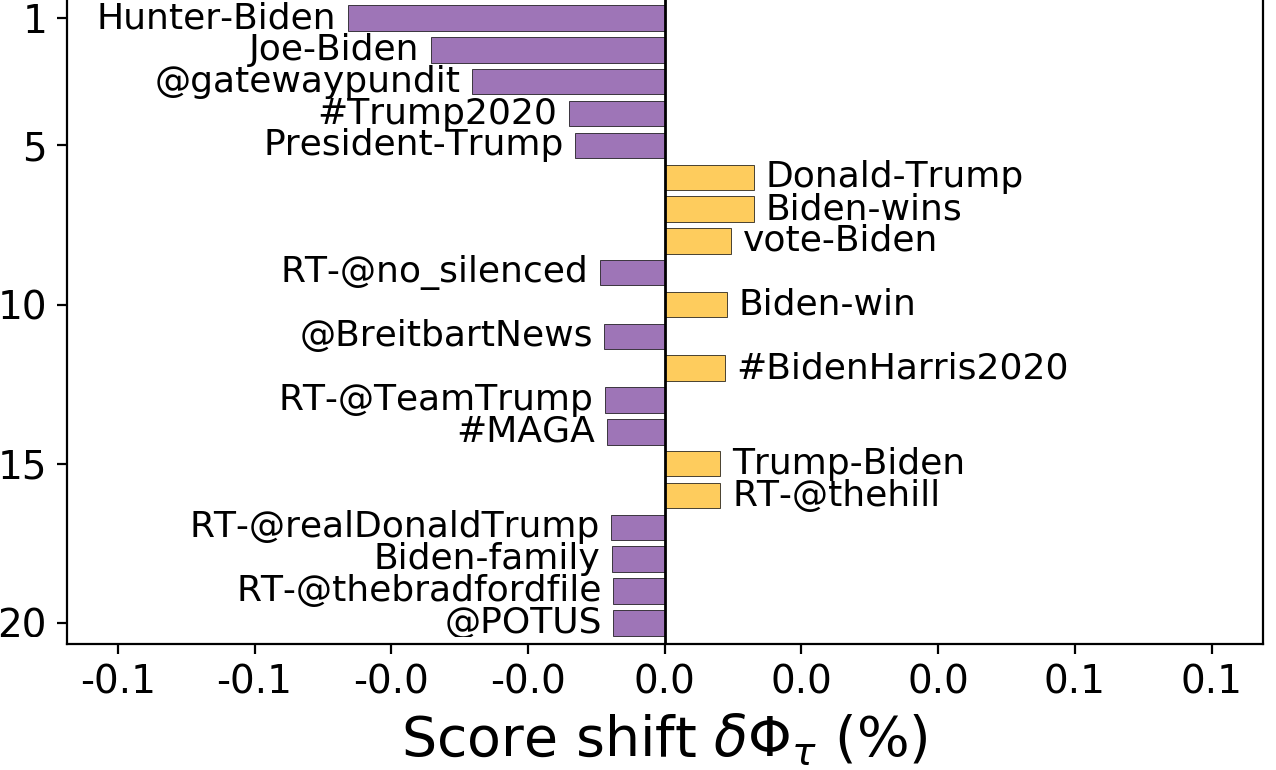}
    \end{subfigure}
    \begin{subfigure}{.20\textwidth}
        \centering
        \small Brexit (2015)
        \includegraphics[width=\linewidth]{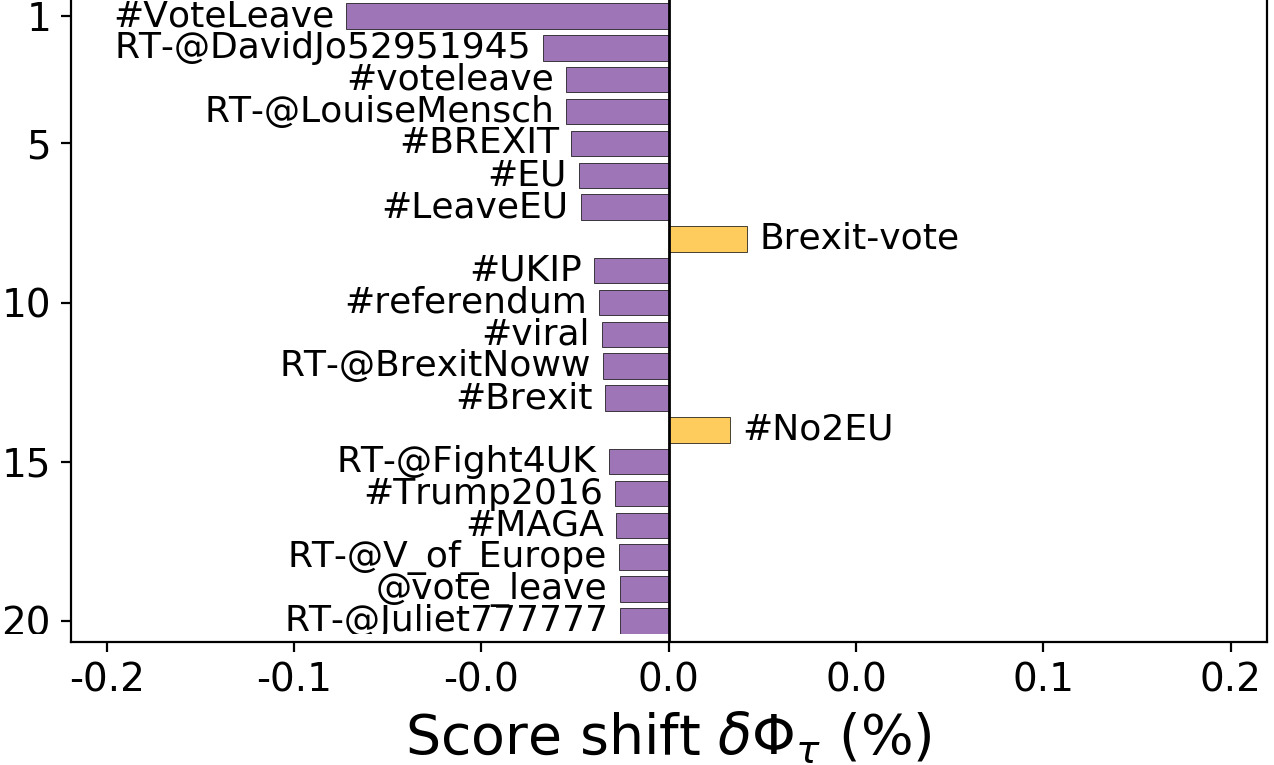}
    \end{subfigure}
    \begin{subfigure}{.20\textwidth}
        \centering
         \small Christchurch (2019)
        \includegraphics[width=\linewidth]{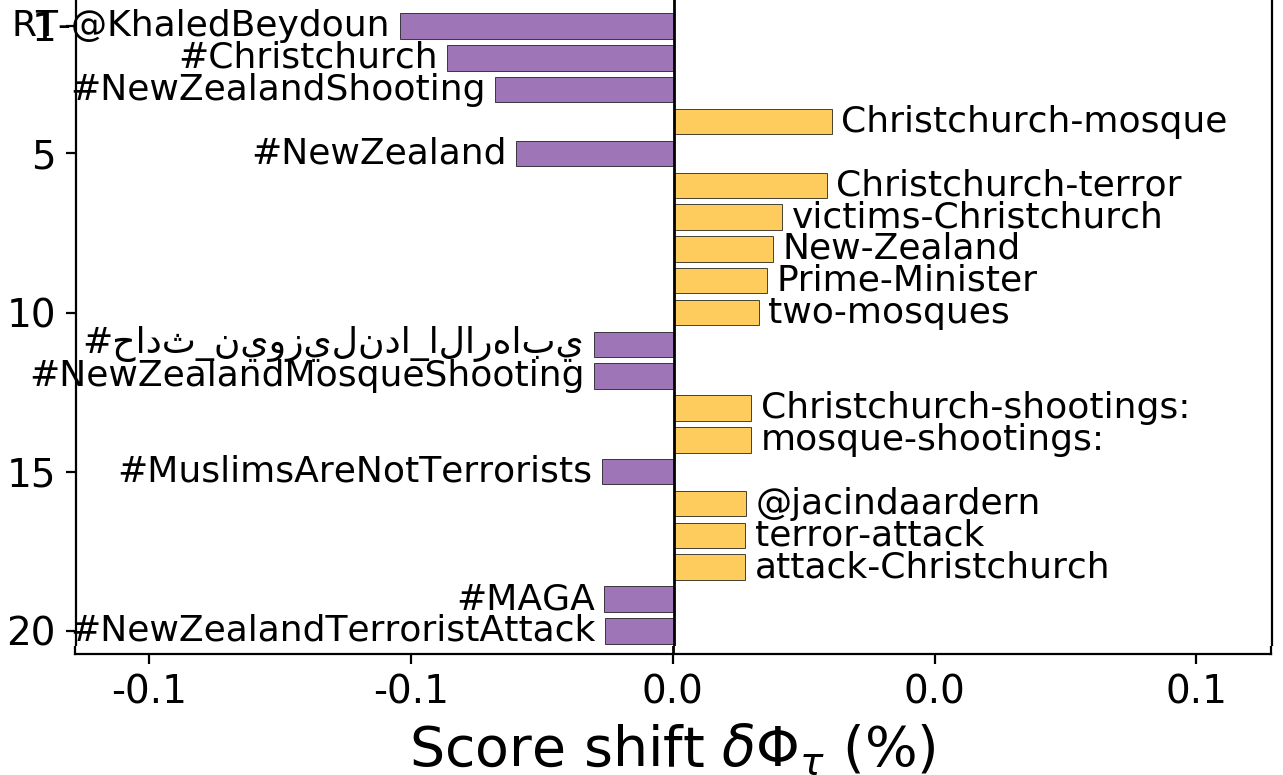}
    \end{subfigure}
    \begin{subfigure}{.19\textwidth}
        \centering
         \small Gun Control (2018)
        \includegraphics[width=\linewidth]{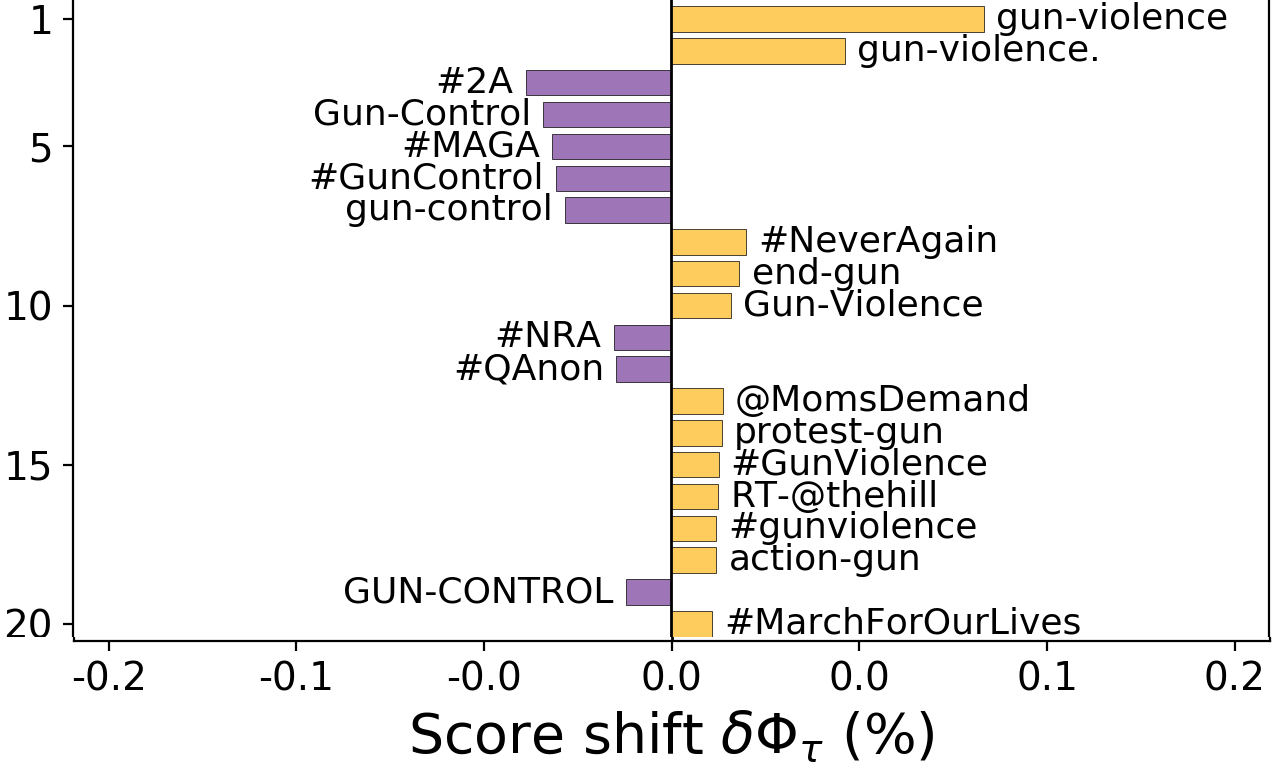}
    \end{subfigure}
    \begin{subfigure}{.19\textwidth}
        \centering
         \small Hillary (2016)
        \includegraphics[width=\linewidth]{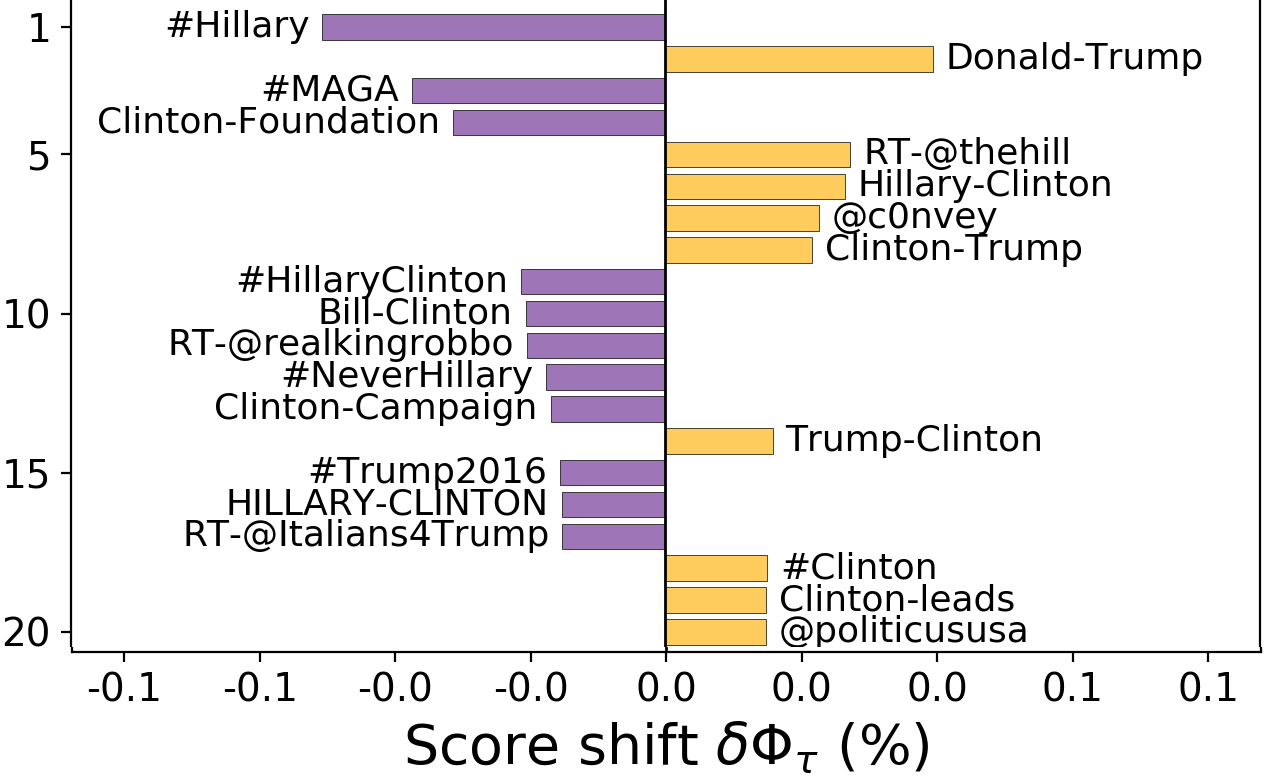}
    \end{subfigure}
    \\
    \begin{subfigure}{.20\textwidth}
        \centering
        \small Kavanaugh (2018)
        \includegraphics[width=\linewidth]{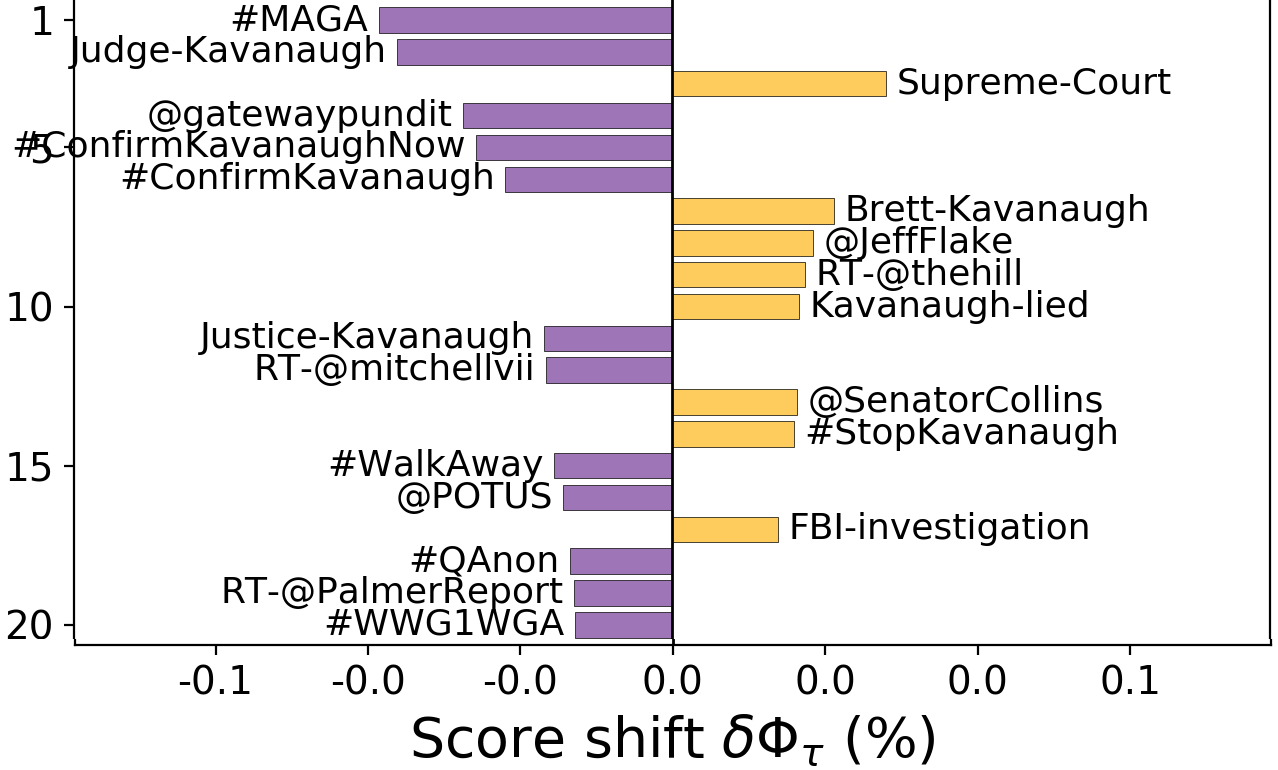}
    \end{subfigure}\hfill
    \begin{subfigure}{.20\textwidth}
        \centering
        \small Macron (2017)
        \includegraphics[width=\linewidth]{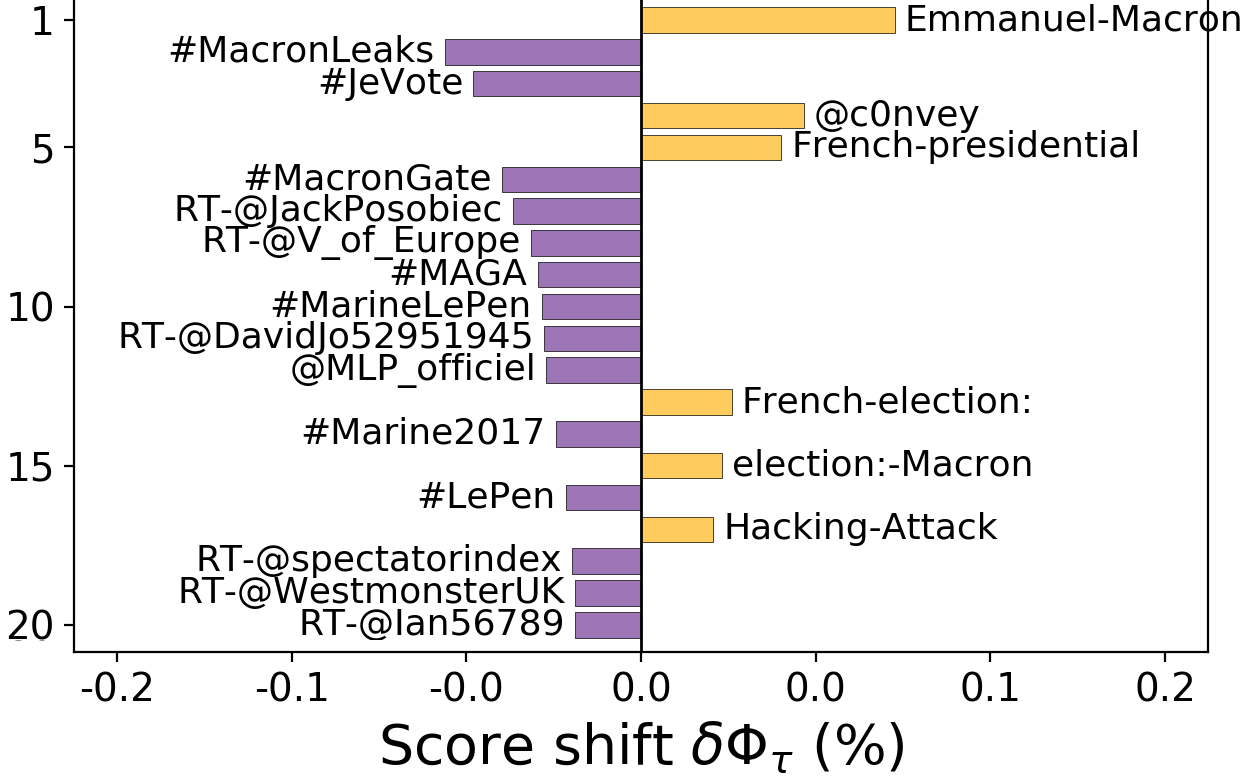}
    \end{subfigure}
    \begin{subfigure}{.20\textwidth}
        \centering
         \small Netflix (2019)
        \includegraphics[width=\linewidth]{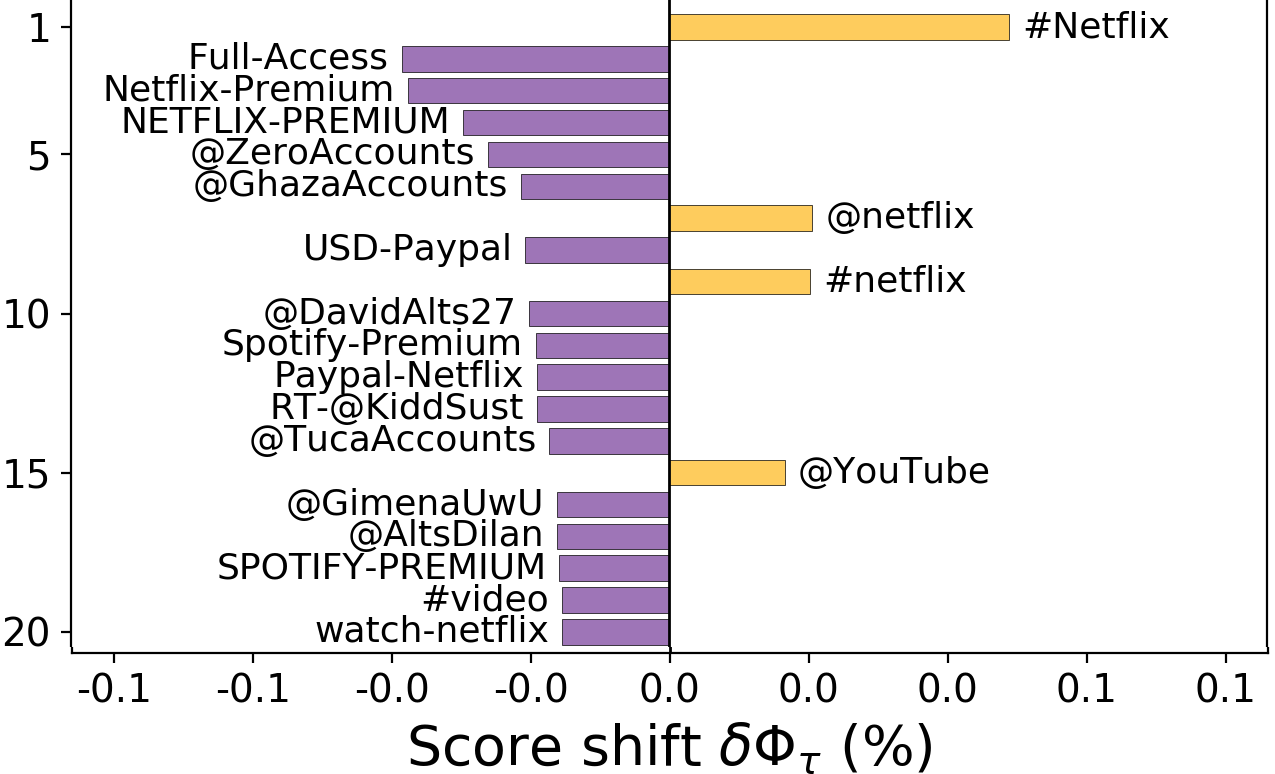}
    \end{subfigure}
    \begin{subfigure}{.19\textwidth}
        \centering
         \small Obama (2015)
        \includegraphics[width=\linewidth]{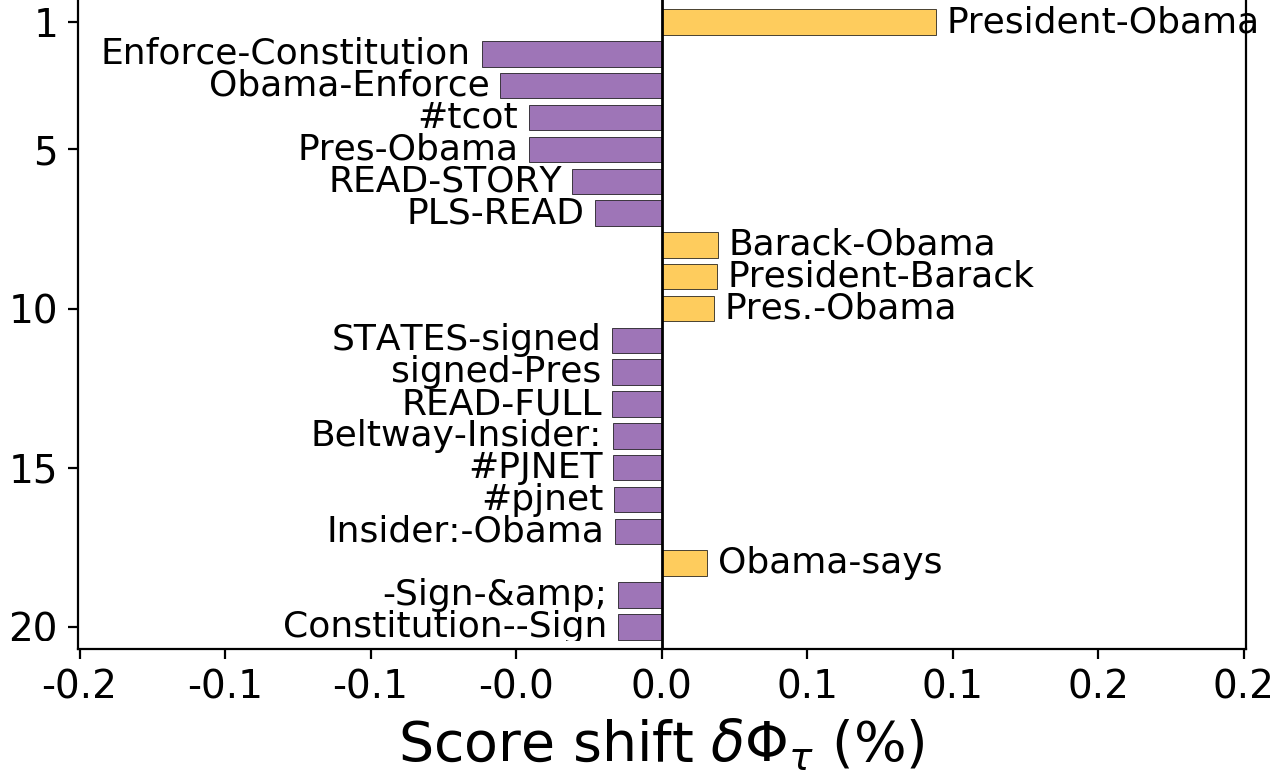}
    \end{subfigure}
    \begin{subfigure}{.19\textwidth}
        \centering
         \small Paris Agreement (2019)
        \includegraphics[width=\linewidth]{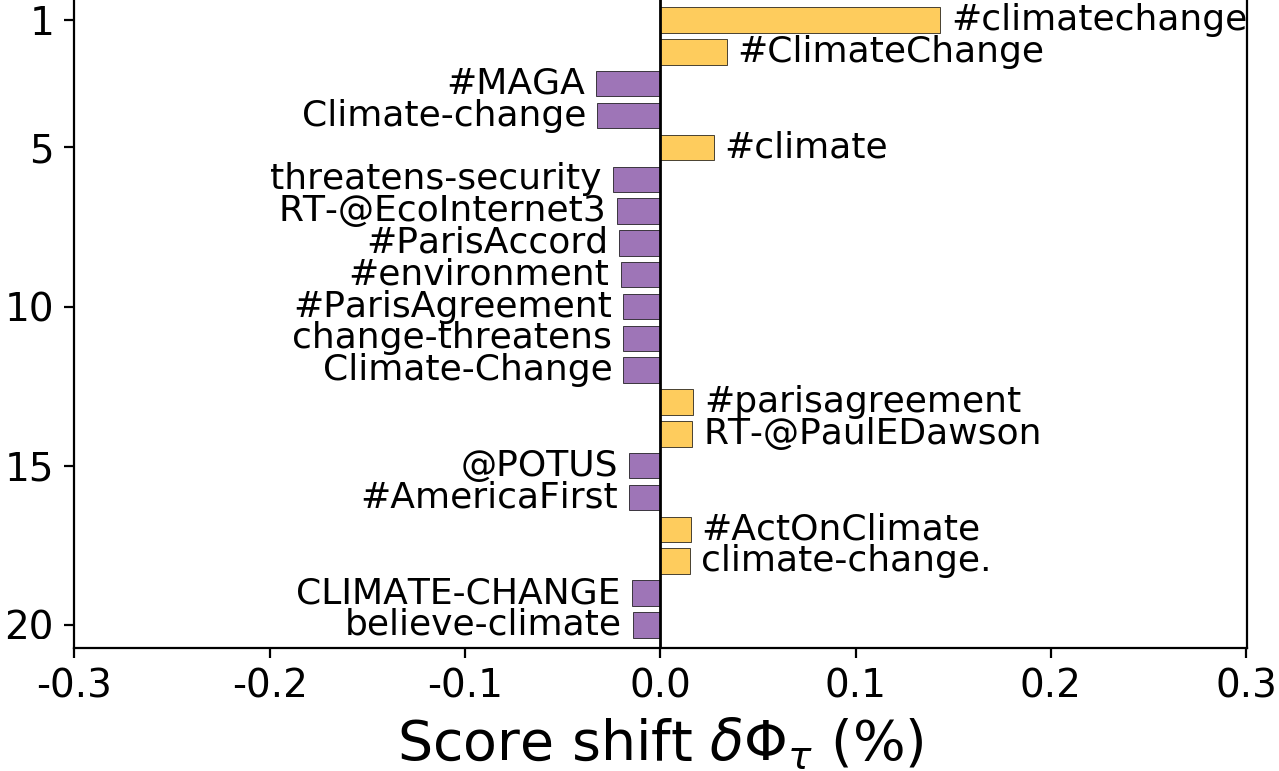}
    \end{subfigure}
    \\
    \begin{subfigure}{.20\textwidth}
        \centering
        \small Pizzagate (2017)
        \includegraphics[width=\linewidth]{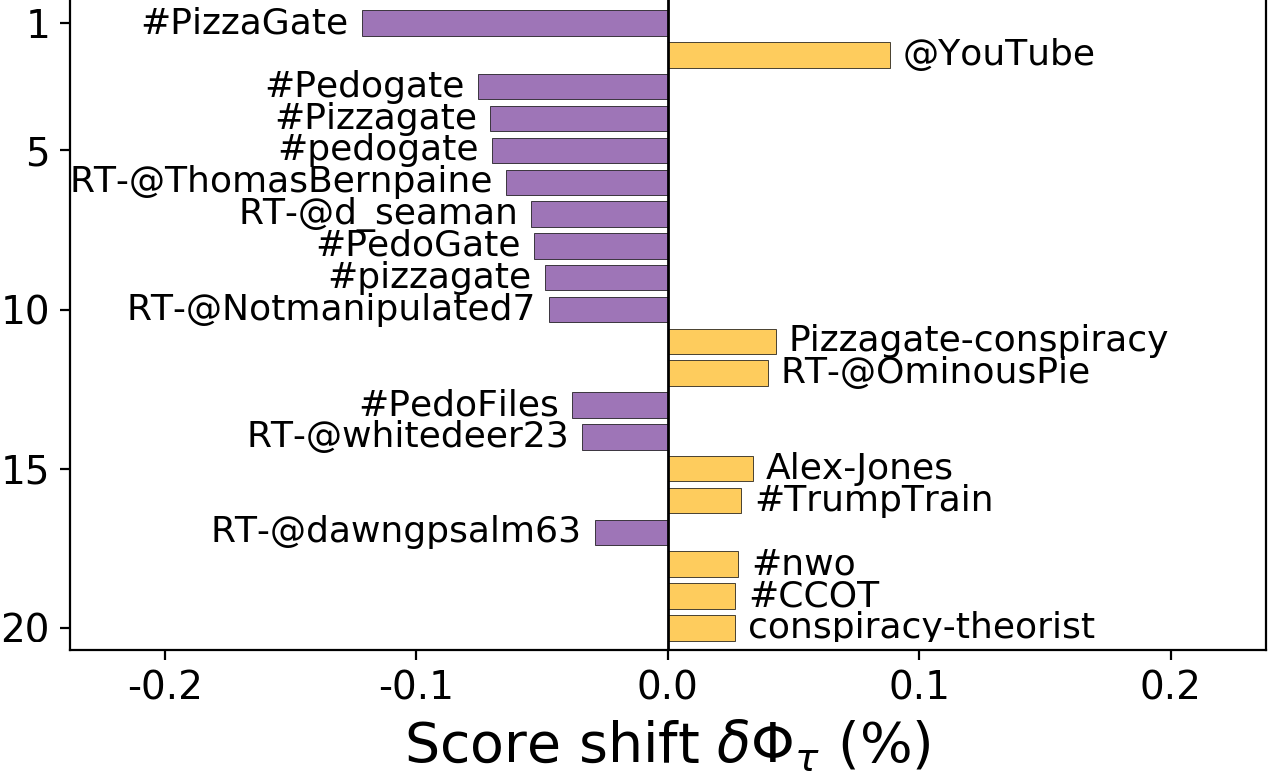}
    \end{subfigure}\hfill
    \begin{subfigure}{.20\textwidth}
        \centering
        \small Pro-Hillary (2016)
        \includegraphics[width=\linewidth]{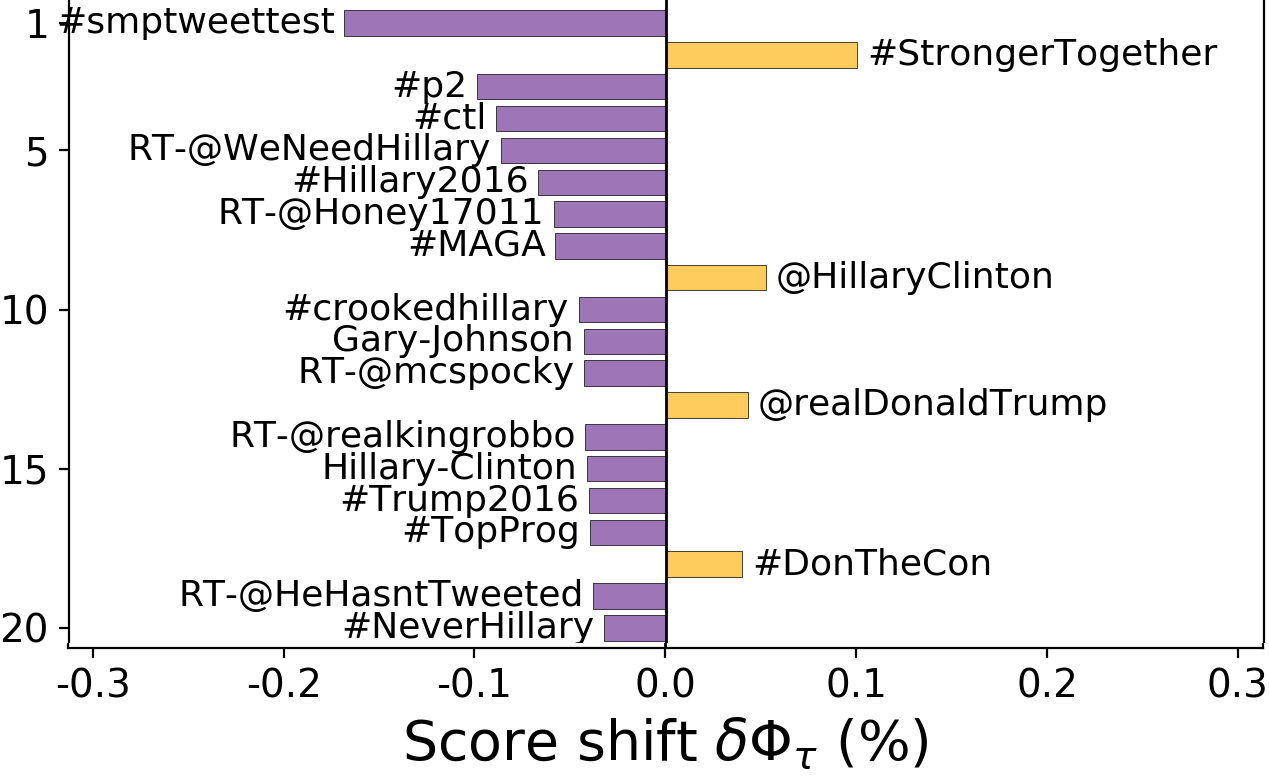}
    \end{subfigure}
    \begin{subfigure}{.20\textwidth}
        \centering
         \small QAnon (2020)
        \includegraphics[width=\linewidth]{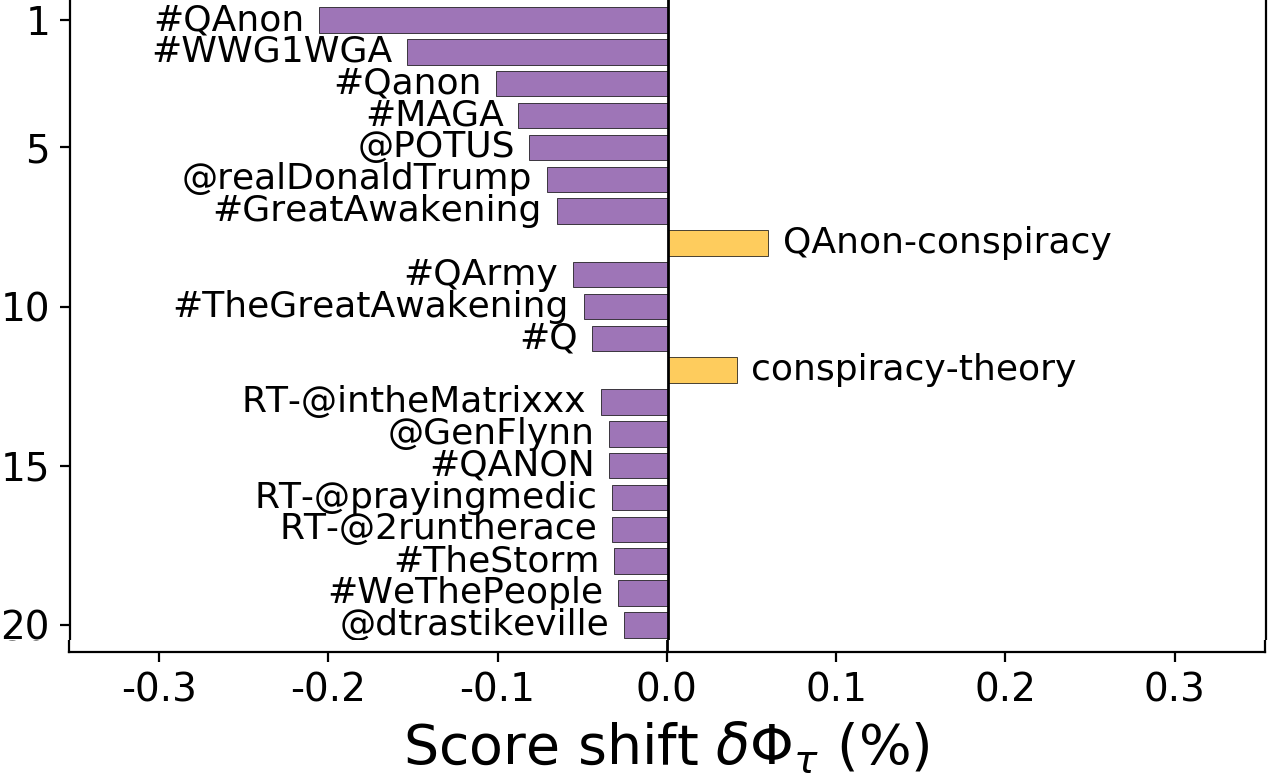}
    \end{subfigure}
    \begin{subfigure}{.19\textwidth}
        \centering
         \small Trump (2020)
        \includegraphics[width=\linewidth]{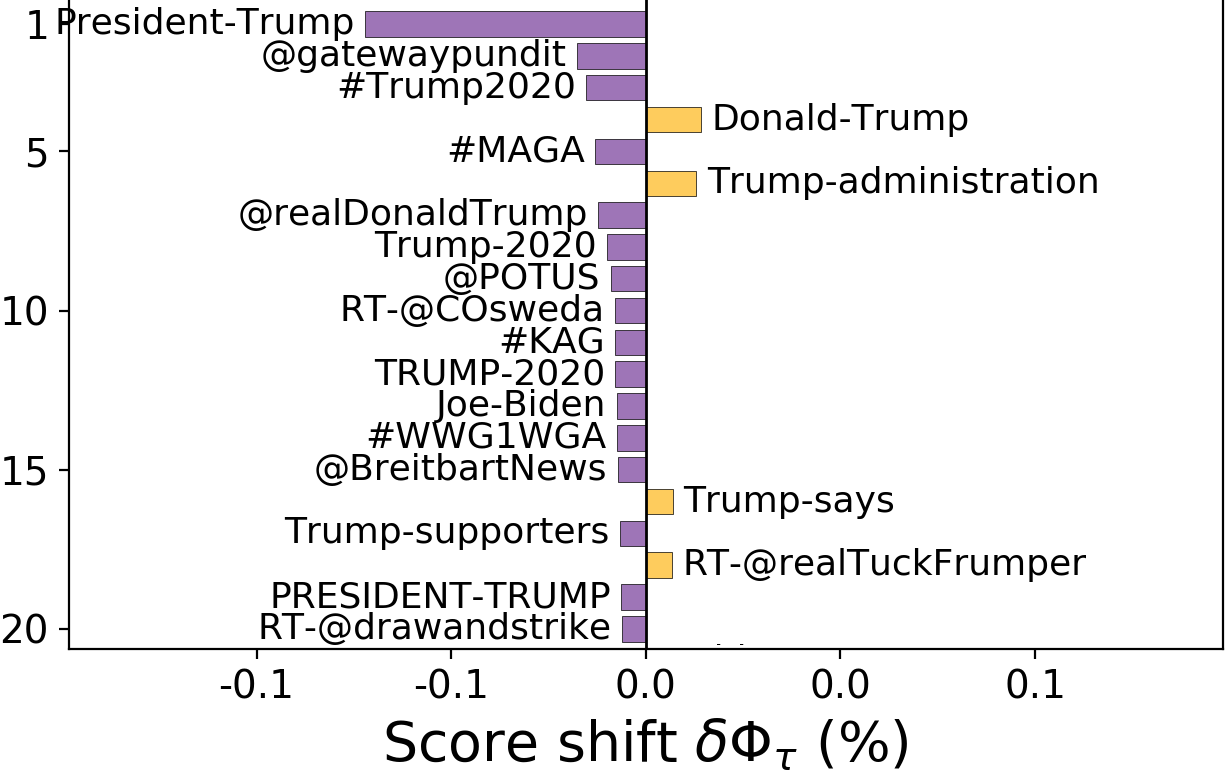}
    \end{subfigure}
    \begin{subfigure}{.19\textwidth}
        \centering
         \small Yellow Vests (2019)
        \includegraphics[width=\linewidth]{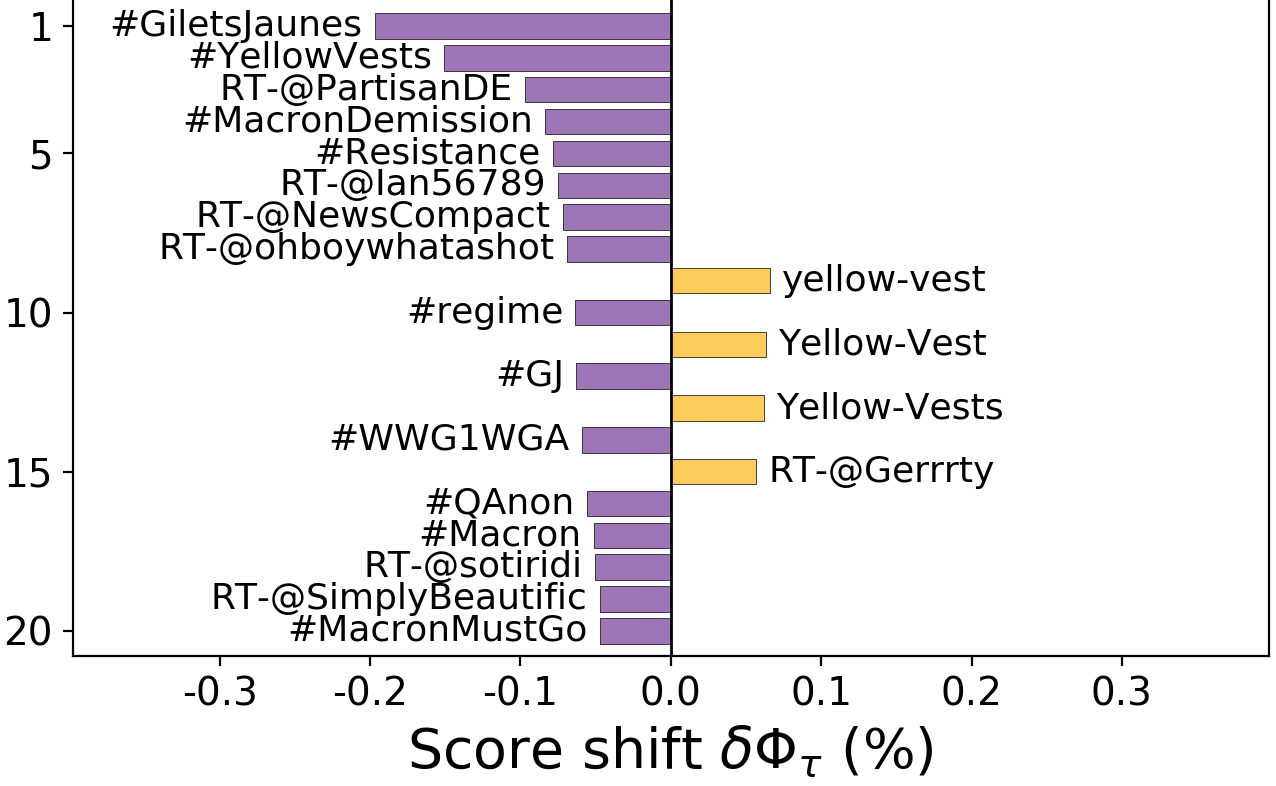}
    \end{subfigure}

    \caption{Word shift graphs depicting the most distinctive keywords of missing tweets (left) and the recollected tweets (right).\label{fig:ex}}

\end{figure*}

\section{Case 2: Trending Topics} 

Twitter trends are the topics that are popular at a moment. Twitter amplifies them to a broader audience by listing them on its user interface. Past work showed that trends are vulnerable to manipulation. In some countries, they are manipulated daily. For instance, Elmas et al.~\cite{elmas2021ephemeral} found that at least 47\% of local trends in Turkey are fake and created from scratch using bots. The authors reported that for 70\% of the fake trends, all of the tweets pushing the trend to the list are deleted, making it impossible to investigate the source of the trend. Correspondingly, they found at least five social media studies attributed a bot campaign to the public as they were not able to collect the data of the bots who were the source of the campaign.

Based on this premise, we analyze if and to what degree the data persistence biases the results of analysis on tweets that originate (i.e. push the keywords on the trends list) the trends. We collect all trends in the United States in 2020. To the best of our knowledge, trends in the United States are not manipulated as regularly as in Turkey. Thus, we assume that the data persistence will be at a regular level. We collect all the tweets mentioning those trends from the dataset provided by the Internet Archive. The initial trend dataset has 62,940 trends. Contrary to the previous case, the trends do not span a long time, and some topics may be popular for less than an hour. Thus, some trends have very few tweets in the 1\% sample. For the most reliable results, we use only the trends with at least 100 tweets in the sample, which results in 13,587 trends.

\begin{figure}[htb]
    \includegraphics[width = 0.85\columnwidth]  {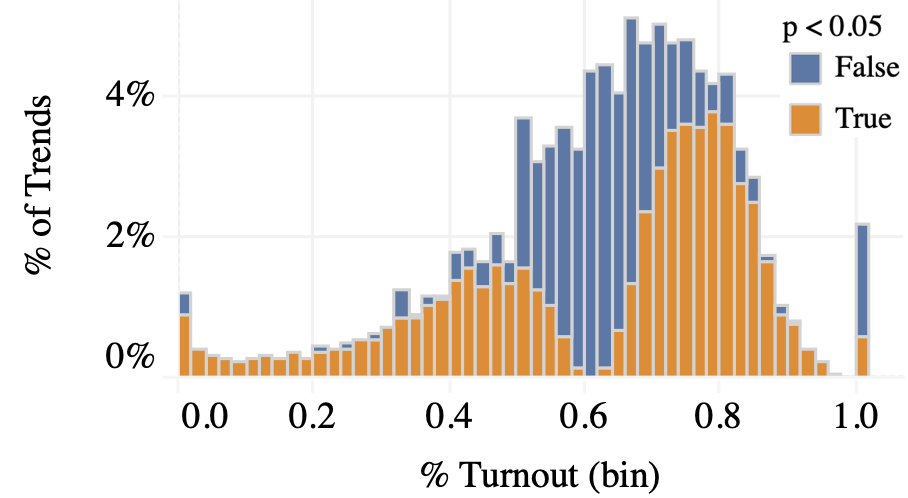}
    \caption{The turnout rates of trends. The average is 62.5\% }
    \label{fig:trend_perc}
\end{figure}
\smallskip
\noindent\textbf{Turnout} \textbf{(RQ1)} We recollect the tweets associated with 13,587 trends and compute the turnout. \Figref{fig:trend_perc} shows the results. We found that the tweets that originate a trend have 62.5\% turnout on average, which is slightly higher than the control dataset. We found that 2942 (21.6\%) of trends have more than 50\% of tweets deleted, meaning that it is impossible to reproduce the majority of the tweets that are source of the trend \textbf{(RQ2)}. The results are significant with p < 0.05 for 2,411 of them (17.7\%). 

We also compared the turnout after the trends appear on the trends list with the turnout before by computing their difference. We find that the difference is low, between -0.05\% and 0.05\% for 41\% of the trends, and the majority of the results (80\%) are not statistically significant (p < 0.05).  

\begin{figure}[htb]
    \includegraphics[width = 0.9\columnwidth]{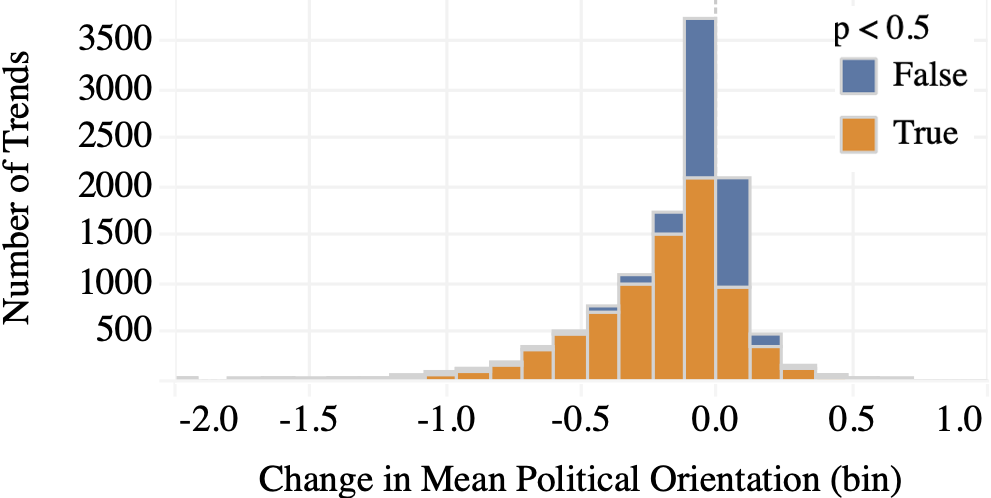}
    \caption{The change of mean political orientation of trends upon recollection. Many are biased toward the left.}
    \label{fig:trends_stance_change}
\end{figure}

\smallskip
\noindent\textbf{Political Orientation} \textbf{(RQ2)} As in the previous case, we found that the mean political orientation of trends changes after recollection. As ~\Figref{fig:trends_stance_change} shows, a significant amount of trends are biased towards left after recollection. Precisely, 1,116 trends are biased toward the right and 4,433 trends are biased toward the left by at least a score of 0.1. However, the impact is low, compared to the previous case. Only 1,998 trends leaned towards the left by a score of 0.4 (the control dataset) and only 395 had more than 1.0 (the results are statistically significant with p < 0.05).

\begin{figure}[htb]
    \includegraphics[width = \columnwidth]{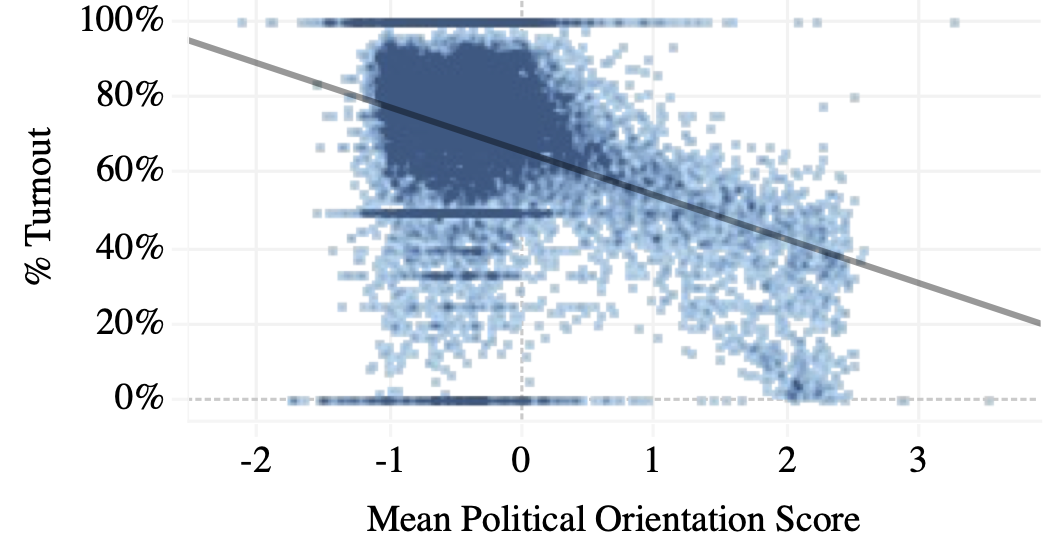}
    \caption{Mean political orientation of the users originating a trend and the trends' turnout. The trends that are originated by right-aligned users are more likely to have lower turnout.}
    \label{fig:scatter_trend_pol}
\end{figure}

Nevertheless, we found that the trends that are targeted by right-aligned users are more likely to suffer from data persistence \textbf{(RQ3)}. We test this by computing the Pearson correlation between the turnout and the mean political orientation scores of trends. ~\Figref{fig:scatter_trend_pol} shows, the two values are inversely correlated. The Pearson correlation is -0.38 and is statistically significant (p < 0001). 

\begin{figure}[htb]
    \includegraphics[width = 0.7\columnwidth]{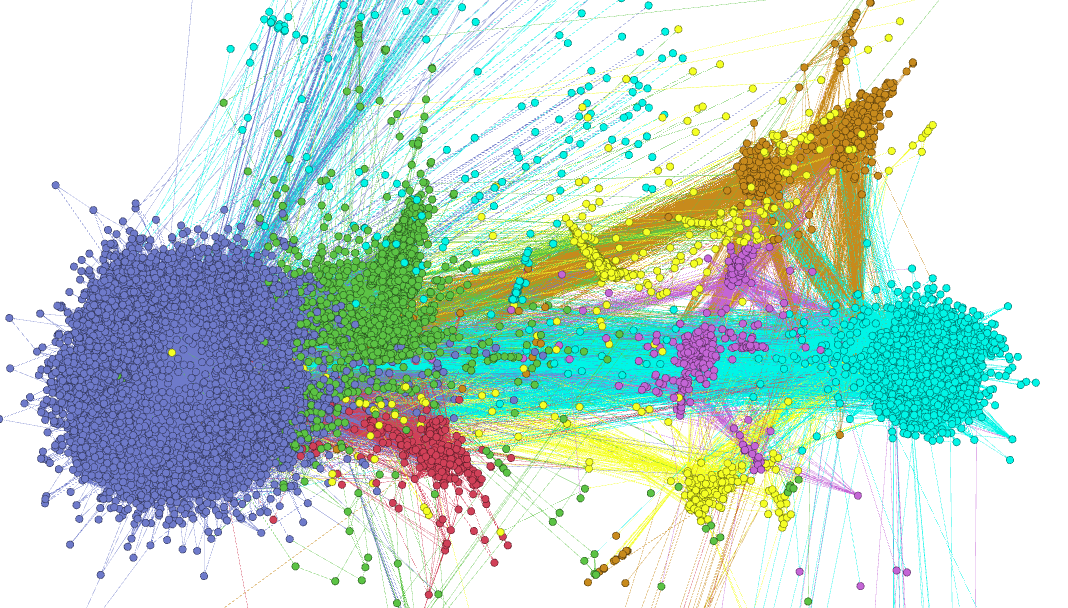}
    \caption{The communities of U.S. trends. The small communities are omitted for visualization purposes.}
    \label{fig:trend_communities}
\end{figure}

\begin{figure}[htb]
    \includegraphics[width = \columnwidth]{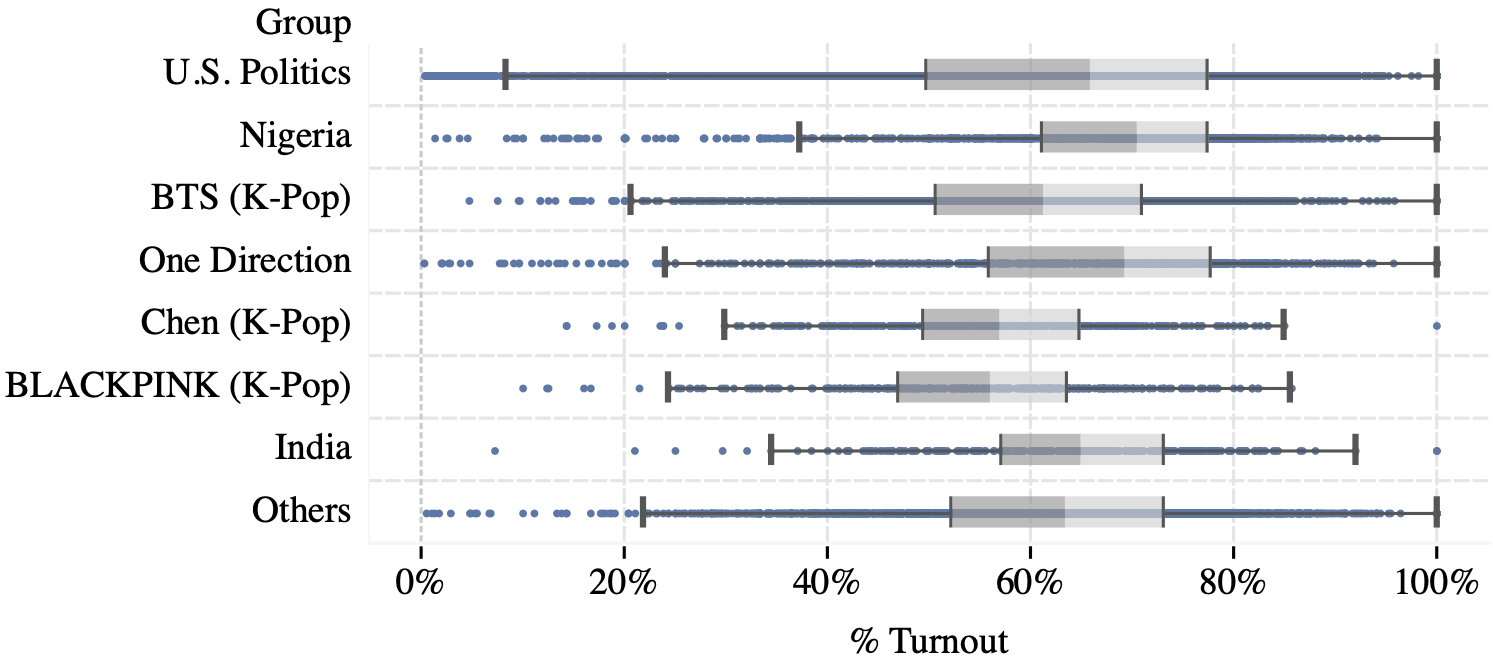}
    \caption{The tweet turnouts for each trend group.}
    \label{fig:trend_turnout}
\end{figure}

\begin{figure}[htb]
    \includegraphics[width = \columnwidth]{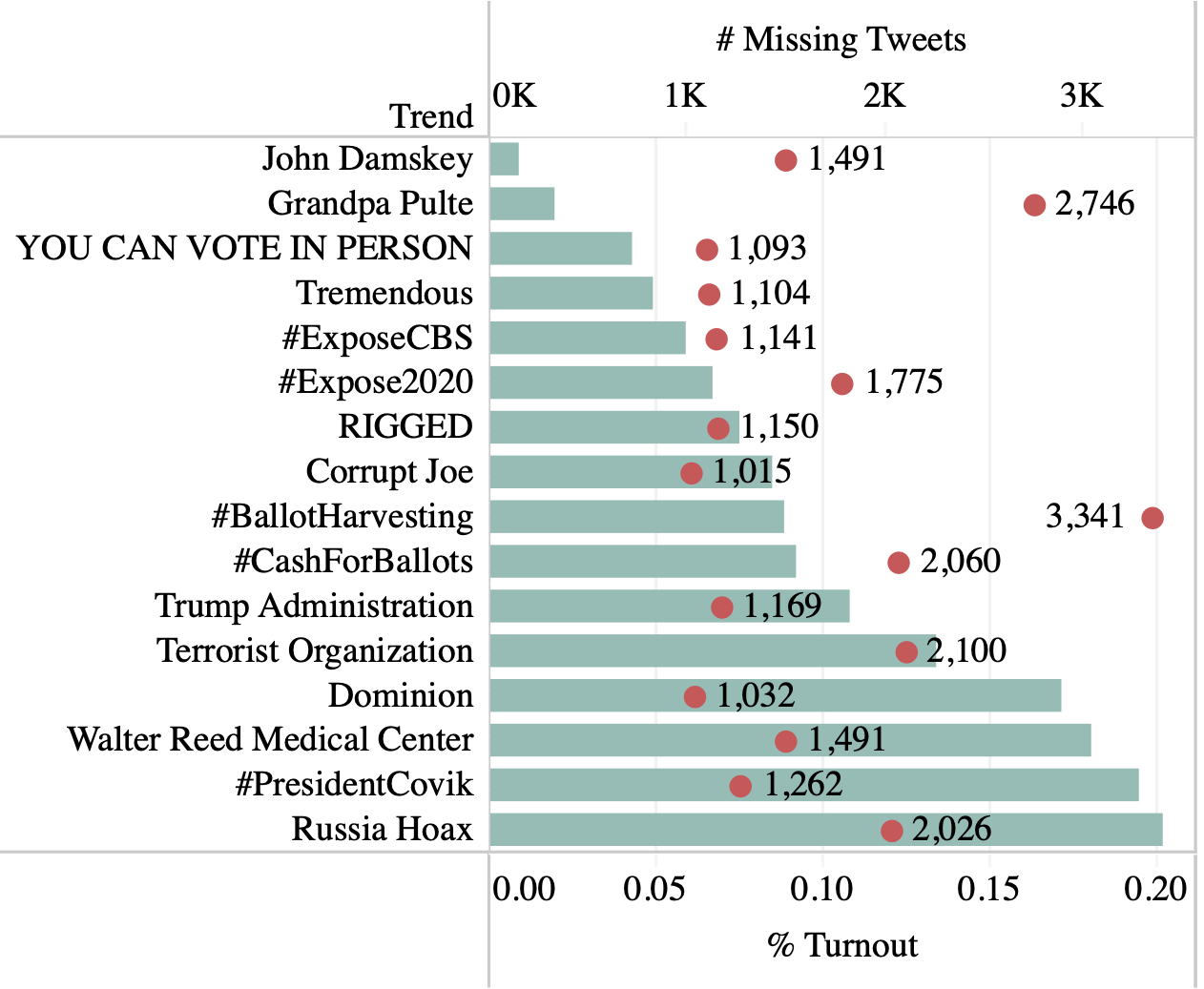}
    \caption{The trends with the least turnout despite being mentioned by at least 1000 tweets in the 1\% sample. Most are related to U.S. politics and some involve manipulation.}
    \label{fig:least_trends}
\end{figure}

\smallskip
\noindent\textbf{Topics} To answer \textbf{RQ3}, we categorized the trends into topics. As the data is on a larger scale, we opted out of employing a textual analysis and used a network-based approach instead. We created a network of trends where nodes are trends and an edge between two trends indicates that they have a common user base, i.e. the edge weights denote the number of users that posted to both trends. To have well-defined trend groups, we only kept the edges with a weight of at least 10. We use Louvain method~\cite{blondel2008fast} to detect the communities which would map to trend groups. ~\Figref{fig:trend_communities} shows the resulting network visualized by Force Atlas 2 by Gephi and colored by the communities. We inspect the nodes with the highest degree in each community to describe the community. We found that the communities map to either the topics related to a specific country or a music group as Table~\ref{tab:communities_table} shows. We observe that the community related to Nigeria had some noisy trends such as "Kobe", but the trends were overwhelmingly related to Nigeria and "End Sars" protests. Next, we compute the turnout for the trends in each group. ~\Figref{fig:trend_turnout} shows the results. We found that the trends related to K-Pop groups are more likely to have less turnout. BTS-related trends have a median of 61\% turnout, Chen 57\%, and Black-Pink 56\%. Trends related to Nigeria have more turnout than U.S. and India, with a median turnout of 71\%. It is unclear if this is because there is a less malicious activity in Nigeria or the effort in content moderation is low. Meanwhile, the trends related to U.S. politics have a very low lower quartile (50\%) and lowest lower whisker among all groups (8\%). In fact, there are many trends that do not have any data left to analyze. \Figref{fig:least_trends} shows the trends with the least turnout despite having at least 1000 tweets in the 1\% sample. All these trends are related to U.S. politics and some appear to be related to political manipulation such as \#BallotHarvesting.

\begin{table}[]
\caption{The communities and the top trends in each of them.}
\label{tab:communities_table}
\begin{tabular}{p{50pt}|p{20pt}|p{140pt}}
\hline
Name                & Color  & Top Trends (Sorted By Degree)                                                                                                                     \\ \hline
U.S. Politics       & Blue   & \begin{tabular}[c]{@{}l@{}}ally, Iran, Donald, \\ Justice, Roger, \#VPDebate\end{tabular}                                                         \\ \hline
Nigeria             & Green  & \begin{tabular}[c]{@{}l@{}}Kobe, \#EndSAR, \#SARSMUSTEND\\ Lekki, ether, rema\end{tabular}                                                        \\ \hline
BTS (K-Pop)         & Cyan   & \begin{tabular}[c]{@{}l@{}}AMAs, \#BANGBANGCON\_D2, \\Jungkook, \#StayGoldMV,\\ \#ExaBFF, \#StayGold\end{tabular}                                    \\ \hline
One Direction (Pop) & Yellow & \begin{tabular}[c]{@{}l@{}}\#10YearsOfOneDirection, \#10YearsOf1D\\ \#1DOnlineConcertStayAtHome, \\BETTER BY ZAYN,  \#BLUEFALL, Golden\end{tabular} \\ \hline
Chen (K-Pop)        & Brown  & \begin{tabular}[c]{@{}l@{}}Cong, Chen, Jongdae, Chan\\ \#CHEN\_STAYS, \#MyAnswerIsEXO9\end{tabular}                                               \\ \hline
BLACKPINK \newline(K-Pop)         & Pink   & \begin{tabular}[c]{@{}l@{}}VMAs, How You Like That, \\\#LovesickGirls, BLACKPINK\\ blackpink, \#LalisaManobanDay\end{tabular}                       \\ \hline
India               & Red    & \begin{tabular}[c]{@{}l@{}}Arnab, \#SushantSinghRajput, \\\#ArnabGoswami\\ \#Hathras, Kargil, \#RishiKapoor\end{tabular}                            \\ \hline
\end{tabular}
\end{table}


\section{Case 3: Framing of Issues}

We finalize our analysis by testing if our findings are consistent with an actual social media study. We reproduce the study by Mendelsohn et al.~\cite{mendelsohn2021modeling} who studied frames on the issue of immigration in the U.S. Frames are aspects of an issue that are emphasized in discussing that issue. Social media studies analyze which frames are prevalent in a discussion to better understand public opinion on the issue. We use them as proxies for different datasets and topics.

The authors of the original study collected immigration-related data through manually selected keywords. The dataset contains 2.6 million tweets from 2018 and 2019. They then annotated a sample of those tweets by the frames they mention. The frames are divided into three categories. Issue-generic frames include frames that emphasize policy-making-oriented aspects such as economics or legal issues. Issue-specific frames focus on how the immigrants are depicted, as heroes, victims, or threats. Other generic frames focus on the text's narrative (thematic or episodic). Refer to the original study for detailed descriptions of the frames.

To collect the data, the authors employed Twitter Decahose API (10\% sample), which provides the data in real time. Thus, the original study does not have an issue with data persistence. However, studies using retrospective collection (i.e., reproducing it) may have such an issue. Data persistence may introduce bias in certain frames and lead to inconsistent results. By reproducing the original study, we better understand the extent and the direction of such bias. 

\begin{figure}[htb]
    \centering
    \includegraphics[width = \columnwidth]{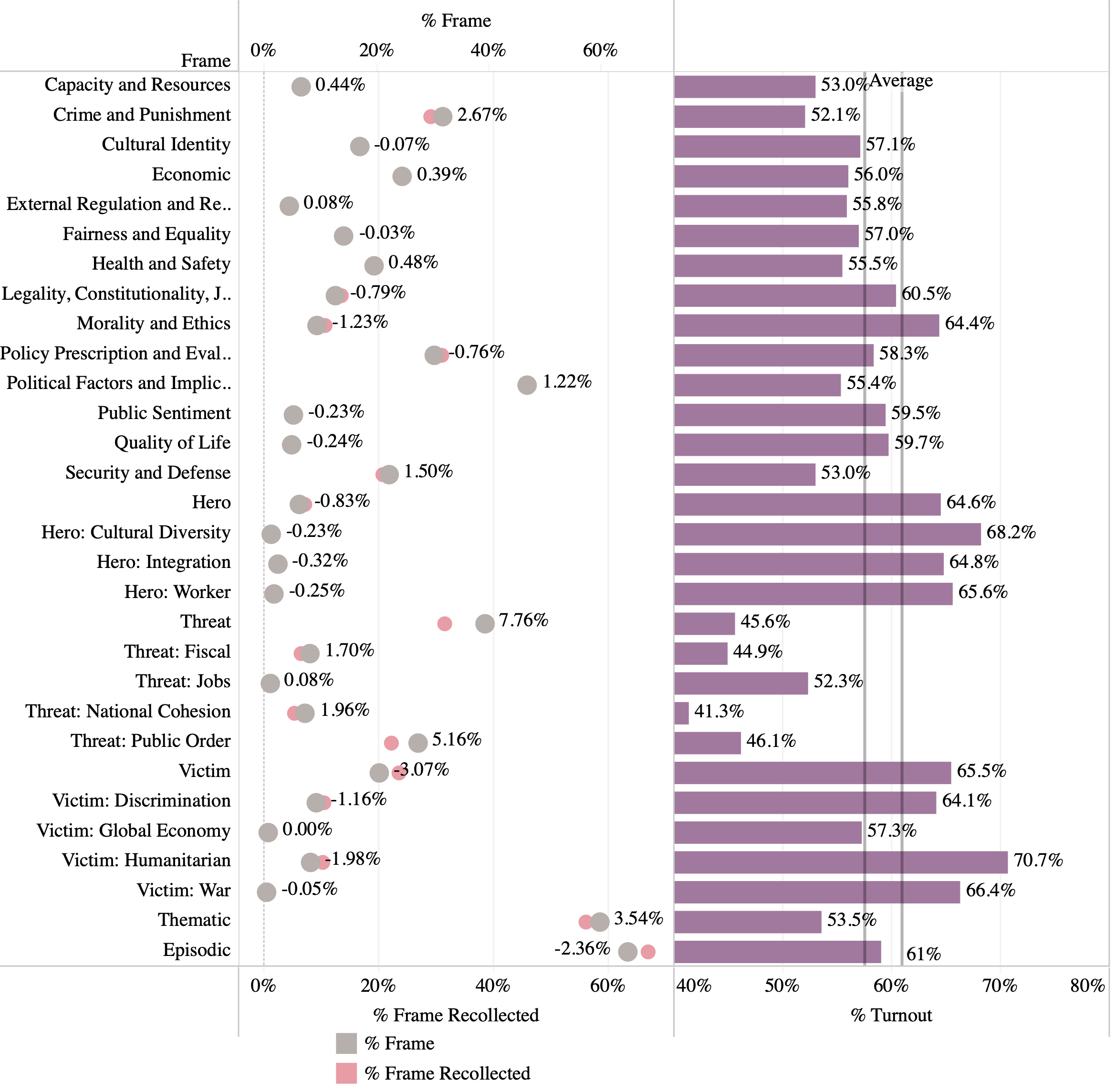}
    \caption{The change in the percentage of frames and their overall turnout. Frames with negative depictions of immigrants are more likely to have low turnout.}
    \label{fig:mendelsohn_turnout}
\end{figure}

\smallskip
\noindent\textbf{Turnout} \textbf{(RQ1, RQ3)} The authors kindly shared their data with us. We recollect their data and compute the turnout, which we found to be 56.9\%, significantly lower than the control dataset (61\%). ~\Figref{fig:mendelsohn_turnout} shows the frame-wise tweet turnouts. The plot on the left indicates the change in the share of the frame in the dataset while the plot on the right indicates the turnout of the tweets mentioning the corresponding frame. We found that the change in the shares is low and the turnouts are close to each other for issue-generic policy frames, although there are still observable differences. The frames that are more likely to mention immigrants in a negative way, Crime \& Punishment (The violation of policies in practice and the consequences of those violations), Capacity \& Resources (The availability or lack of time, physical, human, or financial resources), Security \& Defence (Any threat to a person, group, or nation) has lower turnout, 52\%, 53\%, and 53\% respectively. On the other hand, the frames that may be less likely to have negative depictions such as Morality \& Ethics (Perspectives compelled by religion or secular sense of ethics or social responsibility) and Legality, Constitutionality, Jurisdiction (Court cases and existing laws that regulate policies) have higher turnouts, 64\% and 60\% respectively. The relative ranks of the frames are mostly stable: the order of their prevalence does not change after recollection except that the "Policy Prescription and Evaluation" becomes the second most prevalent frame, over the "Crime and Punishment" after the recollection.

The data turnout is critically low for frames that depict the immigrants as threats, 45.6\%. The turnout of the other issue-specific frames which depict the immigrants as heroes and victims are 64.6\% and 65.5\% respectively, which are higher than the average frame-wise turnout (58\%). The order of the prevalence also changes after recollection, the frame "Victim" becomes more prevalent than "Threat: Public Order" and "Hero" becomes more prevalent than "Threat: Fiscal" and "Threat: National Cohesion." However, the frame "Threat" remains the most prevalent frame.

Thus, we conclude that the data persistence may introduce a bias towards the frames that are positive of the immigrants, and may prevent analyzing the opinions against them. However, it does not introduce a dramatic change to the relative ranks of frames.

\begin{figure}[htb]
    \centering
    \includegraphics[width = \columnwidth]{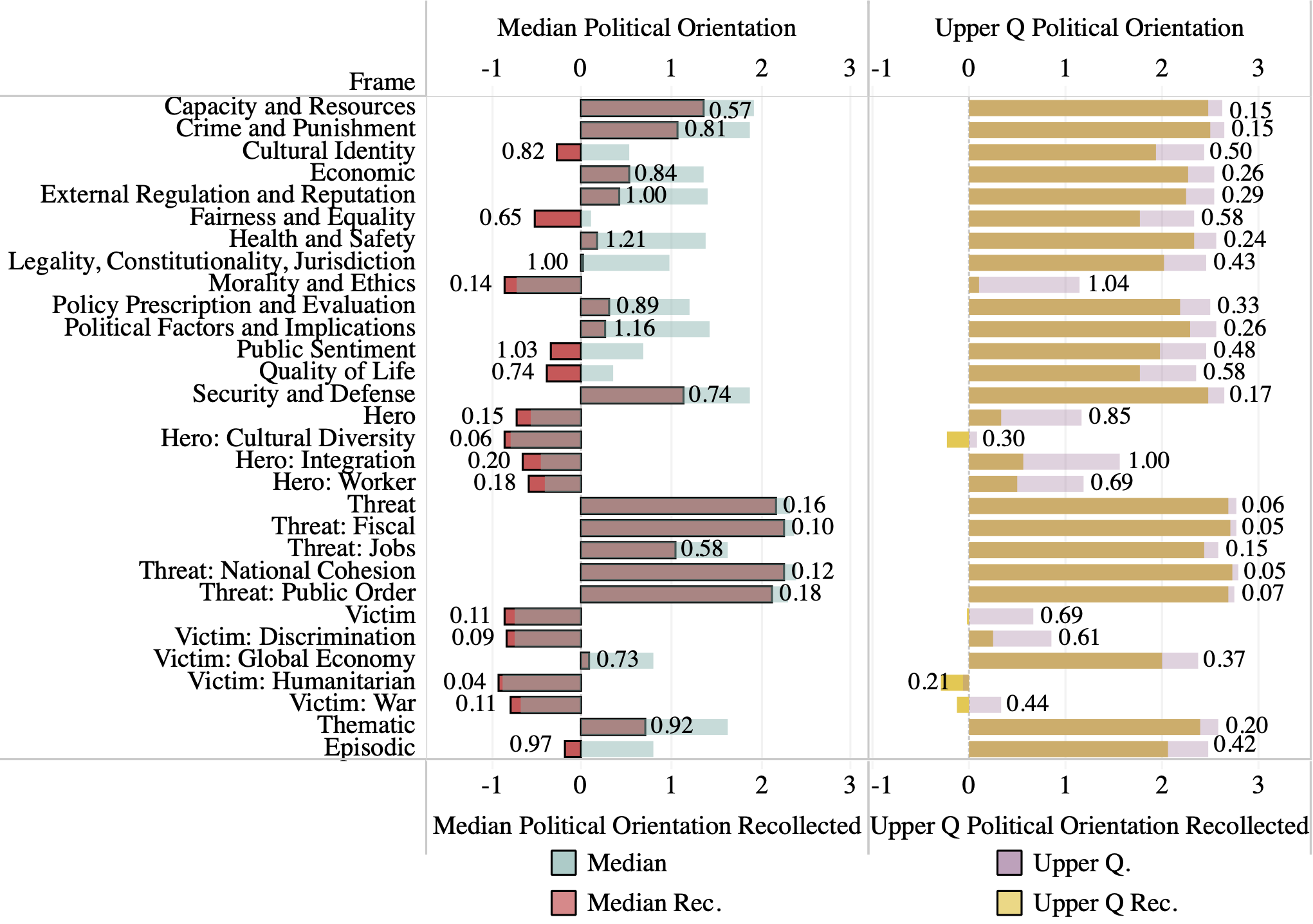}
    \caption{The change in the median and upper quartile of political orientation for each frame.}
    \label{fig:mendelsohn_stance}
\end{figure}

\smallskip
\noindent\textbf{Political Orientation} \textbf{(RQ2)} We also analyzed political orientation with respect to frames. We found that the results are similar to the previous case: almost all the frames (except the ones that are positive towards immigrants and depict them as heroes and victims) are dominated by right-leaning users and the recollection makes the frames lean toward the left. However, we found mixed results when we look at the change in the statistics of the distribution. The magnitude of change in mean political orientation is between 0.16 and 0.45 for all frames, which is lower than the control frames in almost all cases (0.39). However, the change in median political orientation is higher than the control dataset (0.25) for most of the frames, over 1.0 in "External Regulation and Reputation", "Health and Safety" and "Legality, Constitutionality, Jurisdiction", and "Political Factors and Implications". Although the frames that depict immigrants as threats suffered from the data turnout the most, the change in political orientation scores was minimal, and the frames are still largely dominated by right-aligned users. Interestingly, the results are less strong on the upper quartiles than medians contrary to Case 1: none of the frames had more change in the upper quartile than the control dataset (1.3). From these results, we conclude that the data persistence makes datasets lean towards the left more than they actually are, but there are still many right-aligned users whom the researchers can analyze, as the upper quartile did not change as much. For space constraints, we only show the results for the change in medians and upper quartiles in ~\Figref{fig:mendelsohn_stance}.

\section{Factors in Data Removals (RQ4)}

\begin{figure}[!htb]
    \centering
    \includegraphics[width = \columnwidth]{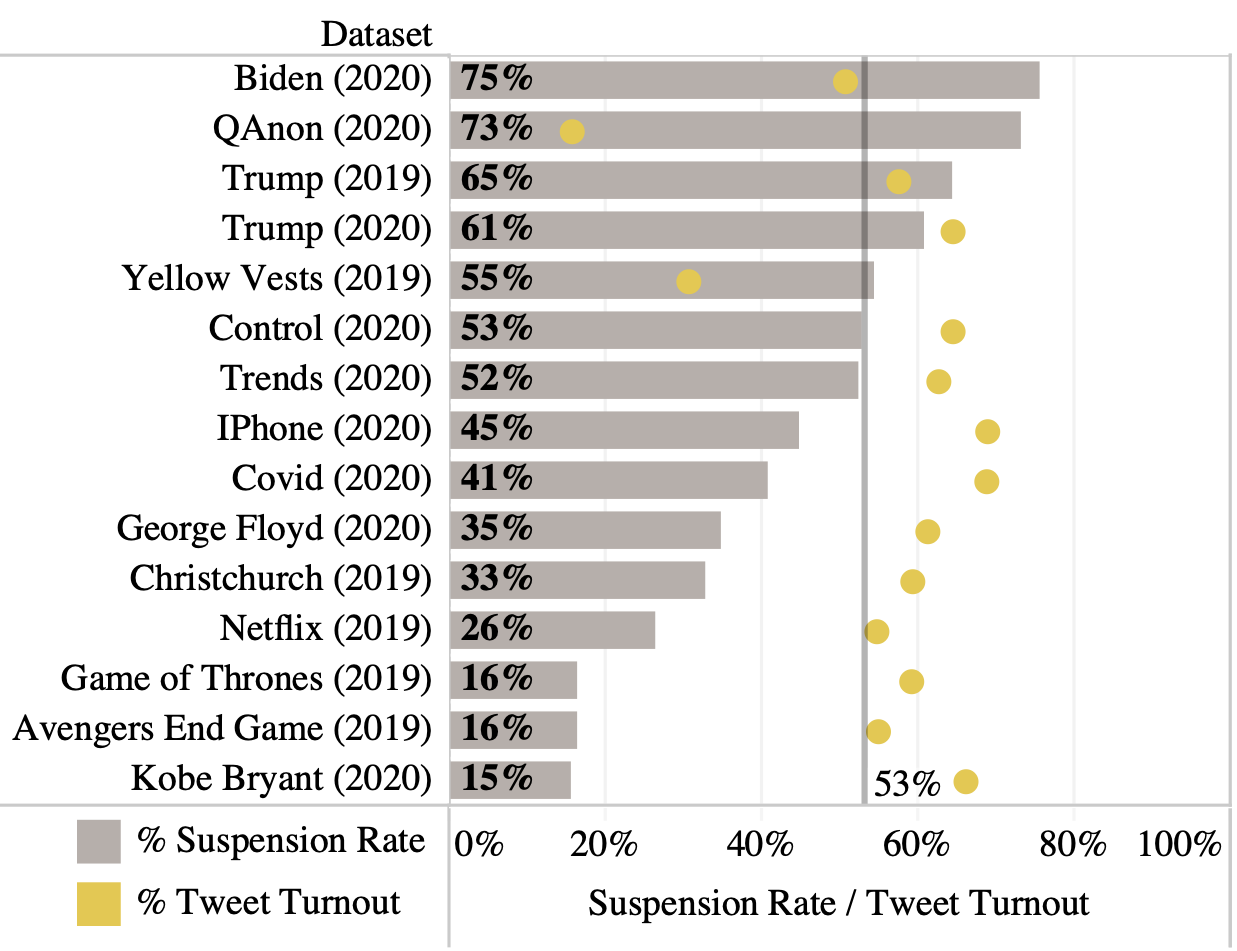}
    \caption{The suspension rates (bars) and the tweet turnouts (dots) for comparison. The vertical line denotes the suspension rate of the control dataset (53\%).}
    \label{fig:suspension}
\end{figure}

The removal of tweets can be due to different factors such as the decision of their authors to remove them or the suspension of their accounts by Twitter. In our analysis, we did not differentiate between these different types of removals as they all reduce data turnout and highlight the need to collect data in real time. However, we hypothesized that controversial topics are more likely to have lower data persistence due to strong content moderation practices resulting in a high rate of account suspension. To investigate this hypothesis, we computed the percentage of uncollectable tweets due to account suspensions using Twitter's new Compliance API endpoint launched in 2021~\cite{compliance}. The endpoint reports which tweets are uncollectible and the reason, including suspension, deletion, account deactivation, or switching to protected mode at the time. Due to potential technical problems on Twitter's side, the endpoint did not return the suspensions for the datasets prior to 2019. Thus, we only report the results for the datasets from 2019 and onwards.

Our results show that in general, roughly half of the uncollectable tweets are due to suspensions, which is the case with the tweets in the control dataset (53\%) and the trends dataset (52\%) as \Figref{fig:suspension} shows. Meanwhile, we observe higher suspension rates for more controversial topics Biden, QAnon, Trump, and Yellow Vests datasets. On the other hand, the suspension rates are very low for non-controversial topics such as Netflix (26\%), Game of Thrones (16\%), Avengers End Game (16\%), and Kobe Bryant (15\%). The Pearson correlation coefficient between the suspension rate and the tweet turnout is -0.42, indicating an inverse correlation. However, the correlation is not statistically significant (p = 0.11) which signifies the need for more datasets to verify the finding. 

Overall, account suspension is the biggest factor in tweet removals as 54\% of the tweets are uncollectable due to it across all datasets. The second biggest factor is deletions, which account for 34\% of the removals. Removals due to accounts being protected make up 10\%. However, the impact of such accounts may be ephemeral as accounts may switch back and forth between protected status to public status.
\section{Conclusion} 

In this paper, we show the impact of data persistence by quantifying the bias it introduces using three case studies. We observe that the data turnout for datasets that focus on controversial topics is low, even if the datasets are recent. The political leanings may differ in the case of retrospective collection, and the researchers may not be able to analyze particular groups such as right-aligned users that are active on the topic reliably. On the other hand, this is less of a concern if the dataset is not on a controversial topic. Additionally, not all groups of interest suffer from data persistence, such as the users who depict immigrants as heroes or victims. We also find that the turnouts of tweets with potentially harmful content are in line with the overall turnout for controversial datasets. The turnout for such content is lower for non-controversial datasets, which may be due to the regular content moderation the platform employs. In light of these findings, we recommend researchers carefully reflect on the nature of the datasets and the groups they will focus on if they are going to collect their data retrospectively. If the dataset or group is likely to suffer from data persistence, the researchers may investigate the potential bias using our methodology, i.e., compare their dataset with a real-time sample such as Twitter's 1\% sample. They may mitigate the data persistence bias by incorporating such real-time data into their study. We also advise them to report the collection time as datasets collected years later than their creation are more likely to suffer from data persistence. Finally, we emphasize the importance of collecting data in real-time and sharing full data with other researchers to prevent data persistence bias. 

\section{Limitations} In this work, we use the 1\% random sample of all tweets provided by Twitter. Past work argued for the reliability of this sample~\cite{morstatter2014biased} for analyzing content (e.g., sentiment analysis), and user activity patterns~\cite{wang2015should}. Some pointed out its drawbacks such as being open to manipulation through automated accounts~\cite{pfeffer2018tampering} and under-representing users that are less active~\cite{gonzalez2014assessing}. We acknowledge the limitations of this sample and the datasets we created based on it. We mitigate them by only working with datasets that are related to major events, which we assume to be less prone to major data manipulations and sampling biases that would affect our results.

We limited our analysis to Twitter as the platform facilitates real-time collection through public API. However, we believe our results may generalize to all social media platforms where the users or platform may delete content retrospectively based on non-uniform decision-making processes. That is, ephemeral content such as stories are all planned to be removed from the platform while posts are not. Selectively removing the latter type of content may introduce data persistence bias on the platform which we analyzed in this study. However, we acknowledge the need for further studies to investigate these biases across different social media platforms and their implications.

\section{Ethics Statement}

Our analysis requires data that is uncollectable through conventional means, i.e. by directly using Twitter's official API. To collect such data, we utilize a dataset that was publicly released by the Internet Archive, which was used extensively by previous work (e.g.,~\cite{tekumalla2020mining}). As the dataset contains content that is removed, it may introduce ethical issues. To mitigate them, we do not inspect the content itself (e.g., annotate it) and do not expose their authors. We only use public tools to analyze the text in the data and only report aggregate statistics. We make the tweet ids publicly available but refrain from sharing their content of them for such ethical reasons.

\bibliographystyle{ACM-Reference-Format}
\bibliography{main}


\end{document}